\documentstyle[12pt,epsf]{article}
\catcode`\@=11
\def \thesection {\arabic{section}.}

\def \be  {\begin{equation}}
\def \ee  {\end{equation}}
\def \ba  {\begin{eqnarray}}
\def \ea  {\end{eqnarray}}
\def \baa {\begin{eqnarray*}}
\def \eaa {\end{eqnarray*}}
\def \bb  {\begin{thebibliography}}
\def \eb  {\end{thebibliography}}
\def \lab #1 {\label{#1}}
\newcommand \ci [1] {\cite{#1}}
\newcommand \bi [1] {\bibitem{#1}}
\newcommand\re[1]{(\ref{#1})}

\def \matrix #1 {\left(\begin{array}{cc} #1 \end{array}\right)}

\def \tr {\mathop{\rm tr}\nolimits}
\def \Im {\mathop{\rm Im}\nolimits}

\def \e  {\mathop{\rm e}\nolimits}

\newcommand\lr[1]{{\left({#1}\right)}}
\newcommand \widebar [1] {\overline{#1}}
\newcommand\bin[2]{\left({#1}\atop{#2}\right)}
\newcommand \vev [1] {\langle{#1}\rangle}

\newcommand \partder [1] {{\partial \over\partial #1}}
\newcommand{\as}{\ifmmode\alpha_{\rm s}\else{$\alpha_{\rm s}$}\fi}

\def \CO {{\cal O}}

\def \CM {{\cal M}}
\def \CK {{\cal K}}
\def \CH {{\cal H}}

\def \CJ {{\cal J}}
\def \CL {{\cal L}}

\font\cmss=cmss12 
\def\inbar{\,\vrule height1.5ex width.4pt depth0pt}
\def\IC{\relax\hbox{$\inbar\kern-.3em{\rm C}$}}
\def\IZ{\relax{\hbox{\cmss Z\kern-.4em Z}}}
\def\IR{{\hbox{{\rm I}\kern-.2em\hbox{\rm R}}}}

\def\IP{{\hbox{{\rm I}\kern-.2em\hbox{\rm P}}}}
\def\II{\hbox{{1}\kern-.25em\hbox{l}}}

\def\numberbysection{\@addtoreset{equation}{section}
                     \def\theequation{\thesection\arabic{equation}}}
\numberbysection

\def\j{\scriptscriptstyle (j)}
\def\U{W}
\def\W{U}
\def\k{\kappa}

\begin{document}

\def\thefootnote{\fnsymbol{footnote}}
\thispagestyle{empty}

\hfill\parbox{50mm}{{\sc LPTHE--Orsay--97--13} \par
                         hep-th/9704079        \par
                         March, 1997}
\vspace*{35mm}
\begin{center}
{\LARGE Solitons in high-energy QCD}
\par\vspace*{15mm}\par
{\large G.~P.~Korchemsky}$\ {}^{1,2}$
and
{\large I.~M.~Krichever}$\ {}^{3,4}$%
\par\bigskip\par\medskip
${}^1$
{\em Laboratoire de Physique Th\'eorique et Hautes Energies%
\footnote{Laboratoire associ\'e au Centre National de la Recherche
Scientifique (URA D063)} \\
Universit\'e de Paris XI, Centre d'Orsay, b\^at. 211\\
91405 Orsay C\'edex, France}
\par\medskip
${}^2${\em Laboratory of Theoretical Physics,\\
Joint Institute for Nuclear Research, \\
141980 Dubna, Russia}
\par\medskip
${}^3${\em Department of Mathematics, \\
Columbia University, U.S.A.}
\par\medskip
${}^4${\em Landau Institute for Theoretical Physics, \\
Kosygina str. 2, 117940 Moscow, Russia}
\end{center}

\vspace*{20mm}

\begin{abstract}
We study the asymptotic solutions of the Schr\"odinger equation
for the color-singlet reggeon compound states in multi-color QCD.
We show that in the leading order of asymptotic expansion,
quasiclassical reggeon trajectories have a form of the soliton waves
propagating on the 2-dimensional plane of transverse coordinates.
Applying the methods of the finite-gap theory we construct their
explicit form in terms of Riemann theta-functions and examine their
properties.
\end{abstract}

\newpage

\def\thefootnote{\arabic{footnote}}
\setcounter{footnote} 0

\section{Introduction}

It widely believed that the Regge asymptotics of hadronic scattering
amplitudes in high-energy QCD should be described by an effective Regge
theory. In this effective theory reggeized gluons, or reggeons,
play the role of a new collective degrees of freedom. Reggeons form a color
singlet compound states, QCD Pomerons and Odderons, which propagate in
the $t-$channel between scattering hadrons and contribute to a power
rise with energy of the physical cross-sections \ci{Col}. In the
Bartels-Kwiecinski-Praszalowicz approach (BKP) \ci{BKP}, the color
singlet reggeon compound states are built from a conserved number
$N=2,3,...$ of reggeons. For a given number of reggeons, $N$,
the wave function of these states $\chi_N$ depends only on the
transverse reggeon coordinates $b_{\perp,j}=(x_j,y_j)$ ($j=1,...,N$)
and it satisfies the BKP equation
\be
{\cal H}_N |\chi_N \rangle = \varepsilon_N |\chi_N \rangle
\lab{BKP}
\ee
where ${\cal H}_N$ is an effective QCD Hamiltonian acting
on the transverse coordinates of $N$ interacting reggeons. QCD
Pomeron and Odderon appear as the states in the spectrum of
${\cal H}_N$ with the maximal energy $\varepsilon_N$.

The BKP equation has a number of remarkable properties in multi-color
limit of QCD, $N_c\to \infty$ and $\as N_c={\rm fixed}$.
Firstly, introducing holomorphic and antiholomorphic coordinates
on 2-dim plane of transverse reggeon coordinates, $z_j=x_j+i y_j$ and
$\bar z_j=x_j - i y_j$, respectively, one can find that the wave
function $\chi_N$ splits into the product of holomorphic
and antiholomorphic parts \ci{Lip2}
$$
\chi_N(x_1,y_1, ..., x_N, y_N; x_0 ,y_0) =
\varphi_N(z_1-z_0,...,z_N-z_0)\cdot
\widebar\varphi_N(\bar z_1-\bar z_0,...,\bar z_N-\bar z_0)\,,
$$
where $(x_0,y_0)$ is the coordinate of the center-of-mass of the
$N$ reggeon compound state. The effective hamiltonian $\CH_N$
is invariant under $SL_2$ transformations
\be
z_j \to \frac{a z_j +b}{c z_j+d}\,,\qquad
\bar z_j \to \frac{\bar a \bar z_j
+\bar b} {\bar c \bar z_j+\bar d}\,,
\qquad
\lab{sl2}
\ee
with $ad-bc=\bar a \bar d -\bar b\bar c=1$ and $j=0,1,...,N$,
while the wave function is transformed
as
\be
\chi_N \to (c z_0+d)^{2h} (\bar c
\bar z_0+\bar d)^{2\bar h} \chi_N\,.
\lab{sl-chi}
\ee
The conformal weights take the values
$h=\frac{1+m}2 - i\nu$ and
$\bar h=\frac{1-m}2 - i\nu$ with
$m=\IZ\,,\ \nu=\IR$
corresponding to the principal value representation of the $SL(2,\IC)$
group. Secondly, the QCD effective Hamiltonian ${\cal H}_N$ is closely
related to the Hamiltonian of XXX Heisenberg magnet of noncompact $SL(2,\IC)$
spin $s=0$ and, as a consequence, the system of $N$ interacting reggeons
possesses a large enough family of conserved charges $q_k$ in order for
the Schr\"odinger equation \re{BKP} to be completely integrable \ci{FK,Lip}.
This implies that instead of solving the BKP equation one can define
the holomorphic wave function $\varphi_N$ as a common eigenfunction of
the family of $N-1$ mutually commuting operators $[q_k,q_n]=0$
\be
q_k=\sum_{1\le j_1 < j_2 < ... < j_k \le
N} z_{j_1j_2} z_{j_2j_3}...z_{j_kj_1} p_{j_1} p_{j_2}...p_{j_k}\,,
\qquad
k=2,...,N
\lab{q_k}
\ee
with $z_{ab}=z_a-z_b$ and $p_j=-i\partder{z_j}$. There are
also additional constraints on $\varphi_N$
\be
(L_3+h)\, \varphi_N(z_1,...,z_N) = 0\,,\qquad
L_+\varphi_N(z_1,...,z_N)=0\,,
\lab{con}
\ee
which follow from the transformation properties of the wave function
\re{sl-chi}. Here, the notation was introduced for the $SL_2$ generators
$$
L_3=  i\sum_{k=1}^N z_k p_k\,,\quad
L_+=  i\sum_{k=1}^N z_k^2 p_k\,,\quad
L_-=  -i\sum_{k=1}^N p_k\
$$
with $p_k=-i\partial_k$. The conserved charges $q_k$ defined in
\re{q_k} have the meaning of higher Casimir operators of the $SL(2,\IC)$
group those eigenvalues determine the quantum numbers of the
$N$ reggeon compound states.
In particular,
$$
q_2=-\sum_{j>k} z_{jk}^2 p_j p_k = -L_3L_3-\frac12\lr{L_-L_++L_+L_-}
$$
is a
quadratic Casimir operator and its eigenvalue corresponding to the wave
function $\varphi_N$ is given by $q_2=-h(h-1)$ provided that conditions
\re{con} are fulfilled.

To find the holomorphic reggeon wave
function $\varphi_N$ one has to diagonalize simultaneously the
remaining $N-2$ commuting hamiltonians $q_3$, $...$, $q_N$ on the space
of functions satisfying \re{q_k}. This leads to a system of $N$ coupled
partial differential equations on the wave function $\varphi_N$. The
same system can be also interpreted \ci{Qua}
as a set of Schr\"odinger equations,
in which the conformal weight $h$ plays the role of the effective Planck
constant, $\hbar=1/h$. This suggests to consider the quasiclassical
limit of large $h$ and develop the WKB expansion for the holomorphic
reggeon wave function
\be
\varphi_N(z_1,...,z_N)=\exp( i S_0 + i S_1 + ...)\,,
\lab{WKB}
\ee
where $S_0=\CO(h)$, $S_1=\CO(h^0)$, $...$ at large $h$ and
$S_k=S_k(z_1,...,z_N)=\CO(h^{1-k})$. In a complete analogy with
quantum mechanics,
the leading term, $S_0$, defines the action function for the
system of $N$ reggeons. For a given set of quantum numbers $q_k$,
it satisfies the system of the Hamilton-Jacobi equations
\be
\sum_{k=1}^N z_k \frac{\partial S_0}{\partial z_k}=ih\,,
\qquad
\sum_{k=1}^N z_k^2 \frac{\partial S_0}{\partial z_k}=0\,,
\qquad
\widehat q_\alpha\lr{\vec z, 
\frac{\partial S_0}{\partial \vec z}}=q_\alpha\,,
\qquad \alpha=3,...,N
\,,
\lab{HJ}
\ee
where $\vec z=(z_1,...,z_N)$ and $\widehat q_\alpha(\vec z,\vec p)$ stands
for the symbol of the operator \re{q_k}.

For $N=2$ reggeon state, the BFKL Pomeron, the Hamilton-Jacobi
equations \re{HJ} can be easily solved
$$
S_0(z_1,z_2)=-ih\ln\lr{\frac1{z_1}-\frac1{z_2}}\,,\qquad
\varphi_2(z_1,z_2)=\lr{\frac{z_{12}}{z_1z_2}}^h\times
\left[1+\CO(h^{-1})\right]
\,.
$$
The last expression defines the holomorphic wave function in the
leading order of the WKB expansion. Comparing it with the well-known
expression for the wave function of the BFKL Pomeron \ci{BFKL}
we find that it is exact to all orders in $1/h$.

For $N \ge 3$ reggeon states, the Hamilton-Jacobi equations \re{HJ} can
be solved in the separated variables \ci{pre}. Their solutions describe
the compound $N$ reggeon states as a system of $N$ classical particles
moving along quasiperiodic trajectories. The corresponding action-angle
variables were constructed in \ci{pre}, where the close relation between
classical reggeon trajectories and finite-gap soliton solutions \ci{NMPZ}
was established. The main goal of the present paper is to apply
algebro-geometrical methods \ci{Kr} in order to obtain the explicit form
of the reggeon trajectories on the $2-$dimensional
plane of transverse coordinates in terms of Riemann theta-functions \ci{D}.

\section{Hamiltonian flows}

In the leading order of the WKB expansion, we replace in \re{q_k} the
momenta operators $p_k$ by the corresponding classical functions
to get the family of $N-1$ classical Hamiltonians $q_2$, $...$, $q_N$.
For the system of $N$ reggeons with the coordinates $z_k$, momenta
$p_k$ and the only nontrivial Poisson bracket $\{x_k,p_n\}=\delta_{kn}$,
these hamiltonians generate the hierarchy of the evolution equations
\be
\frac{\partial z_n}{\partial t_\alpha}
=\{z_n,q_\alpha\}=\frac{\partial q_\alpha}
{\partial p_n}\,,
\qquad
\frac{\partial p_n}{\partial t_\alpha}
=\{p_n,q_\alpha\}=-\frac{\partial q_\alpha}
{\partial z_n}\,,
\lab{EE}
\ee
with $t_\alpha$ $(\alpha=2,...,N)$ being the corresponding evolution
times and $\partial_{t_\alpha} q_n
=\{q_n,q_\alpha\}=0$. Their solutions define the reggeon trajectories
$z_k=z_k(t_2,...,t_N)$ subject to the periodicity condition
\be
z_{k+N}(t)=z_k(t)\,,\qquad
p_{k+N}(t)=p_k(t)\,.
\lab{per}
\ee
The system of evolution equations \re{EE} has a sufficient number
of the integrals of motion $q_k$ to be completely integrable. It
admits the Lax pair representation, which can be found using the
equivalence of the system of $N$ reggeons and the XXX Heisenberg
magnet of $SL_2$ spin $0$. Namely, for each reggeon we define the
Lax operator as
\be
L_k=\matrix{1-E z_k p_k & E p_k \\ -E z_k^2 p_k & 1+Ez_k p_k}
=\II+ E\left(1\atop z_k\right)\otimes (-z_k,1)\, p_k\,,
\lab{L_k}
\ee
with $E$ being an arbitrary complex spectral parameter. Then, the
evolution equations \re{EE} are equivalent to the matrix relation
\be
\partial_{t_\alpha} L_k = \{L_k,q_\alpha\}
=A_{k+1}^{(\alpha)} L_k - L_k A_k^{(\alpha)}\,,
\lab{LA}
\ee
where $A_k^{(\alpha)}(E)$ is a $2\times 2$ matrix depending on the
choice of the hamiltonian.
As an example, for the hamiltonian $q_N$ one can obtain
\be
A_k^{(N)}=E\frac{q_N}{z_{k-1,k}}
\bin{1}{z_k} \otimes (z_{k-1}, -1)
\lab{A}
\ee
and the corresponding evolution equations look like
$$
\partial_{t_N} x_k = \frac{q_N}{p_k}\,,\qquad
\partial_{t_N} p_k=q_N\lr{\frac1{z_{k-1,k}}-\frac1{z_{k,k+1}}}\,.
$$
The explicit form of $A_k^{(\alpha)}$ for the remaining hamiltonians
can be deduced from the Yang-Baxter equation for the Lax operator $L_k$ and
it will not be important in the sequel.

The integration of the evolution equations \re{EE} and \re{LA}
is based on the Baker-Akhiezer function $\Psi_k(E;t_2,...,t_N)$. It is defined
as a solution of the following system of matrix relations
\be
L_k (E)\, \Psi_k = \Psi_{k+1}\,,\qquad
\partial_{t_\alpha} \Psi_k = A_k^{(\alpha)}(E)\, \Psi_k\,,
\qquad
\Psi_k=\lr{\phi_k \atop \chi_k}\,,
\lab{BA-def}
\ee
where $\phi_k$ and $\chi_k$ are scalar components. We construct the
monodromy matrix
\be
T(E)=L_N(E) ... L_2(E) L_1(E) = \left({ A(E) \atop C(E)} { B(E)\atop D(E)}
\right)
\lab{T}
\ee
and verify using \re{LA} that its eigenvalues are integrals of motion
due to $\partial_{t_\alpha} T(E) = A_1^{(\alpha)} T - T A_1^{(\alpha)}$.
The monodromy matrix generates the shift $T(E)\Psi_1=\Psi_{N+1}$
and we impose the periodicity condition on the Baker-Akhiezer functions
by requiring $\Psi_k(E)$ to be the Bloch-Floquet function
\be
\Psi_{k+N}(E) = \e^{P(E)} \Psi_k(E)\,.
\lab{Bloch}
\ee
Here, $P(E)$ is an eigenvalue of the monodromy
matrix \re{T} and it satisfies the characteristic equation
\be
\det(T(E)-\e^{P(E)})=\e^{2P(E)}-\Lambda(E)\e^{P(E)}+1=0\,,
\lab{det}
\ee
where $\Lambda(E)=\tr T(E)$ can be calculated using \re{T} and
\re{L_k} to be
a polynomial of degree $N$ in the spectral parameter with the
coefficients given by the integrals of motion
$$
\Lambda(E)= 2+ q_2 E^2 + q_3 E^3 + ... + q_N E^N\,.
$$
Introducing the complex function $y(E)=(\e^{P(E)}-\e^{-P(E)})/E$ one
rewrites \re{det} in the form of algebraic complex curve
\footnote{To make a correspondence with notations of \ci{pre},
one has to redefine the local parameter on the curve as
$E=1/x$.}
\be
\Gamma_N: \qquad
y^2=E^{-2} (\Lambda^2(E)-4)
   =(q_2+q_3E+...+q_NE^{N-2})(4+q_2E^2+q_3E^3+...+q_NE^N)\,.
\lab{curve}
\ee
For any complex $E$ in general position the equation \re{det}
has two solutions for $P(E)$, or equivalently $y(E)$ in \re{curve}.
Under appropriate boundary conditions on $\Psi_k(E)$ (to be discussed
later) each of them defines a branch of the Baker-Akhiezer function,
$\Psi^\pm_k(E)$. Then, being a
double-valued function on the complex $E-$plane, $\Psi_k(E)$ becomes
a single-valued function on the Riemann surface corresponding to the
complex curve $\Gamma_N$. This surface is constructed by gluing
together two copies of the complex $E-$plane along the cuts
$[e_2,e_3]$, $...$, $[e_{2N-2},e_1]$ running between the branching
points $e_j$ of the curve (see for detail Fig.~1), defined as simple
roots of the equation
\be
\Lambda^2(e_j)=4\,,
\qquad
j=1,...,2N-2\,.
\lab{bran}
\ee
Their positions on the complex plane depend on the quantum
numbers $q_2$, $q_3$, .$..$, $q_N$ of the reggeon compound
state. In general, thus defined Riemann surface has a genus $g=N-2$ and
it depends on the number of reggeons, {\it e.g.\/} it is a sphere for
$N=2$ state, or BFKL Pomeron, and a torus for $N=3$ state known as QCD
Odderon.

To distinguish points $Q=Q(E)$ on the Riemann surface lying
above $E$ on the upper and the lower sheets we assign them a sign
$Q^\pm=(\pm,E)$, where the upper (+) sheet corresponds to the
asymptotics $y\sim q_N E^{N-1}$ as $E\to \infty$. In particular,
one finds from \re{det} the behaviour of the Bloch multiplier on
different sheets of $\Gamma_N$ for $E\to\infty$ as
\be
\e^{\pm P(E)} = q_N E^N\times \lr{1+\CO(1/E)}\,,
\lab{Bl}
\ee
where $\pm$ sign corresponds to the choice of the sheet.

\begin{figure}[ht]
\centerline{\epsffile{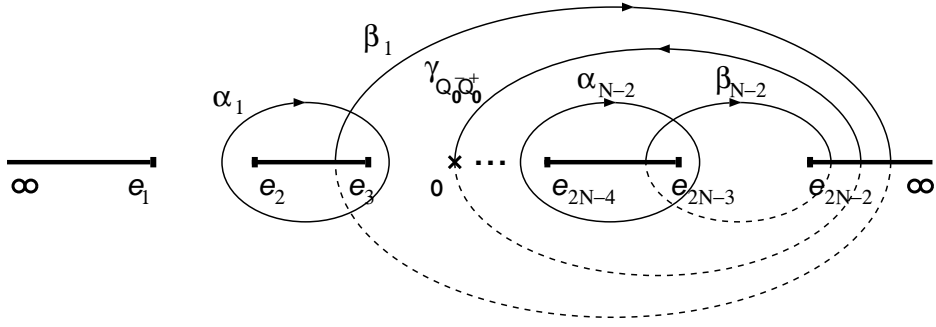}}
\caption{The definition of the canonical basis of cycles on the Riemann
surface $\Gamma_N$. The dotted line represents the part of the $\beta-$cycles
belonging to the lower sheet. Cross denotes a projection of the points
$Q_0^\pm\in \Gamma_N$ on the complex plane. The path
$\gamma_{Q_0^-Q_0^+}$ does not cross the canonical cycles and goes from
the point $Q_0^-$ on the lower sheet to the point $Q_0^+$ on the upper sheet.}
\end{figure}

\subsection{Boundary conditions}

Being written in components, the first condition \re{BA-def} on the
Baker-Akhiezer function looks like
\be
\chi_{n+1}-\chi_n=z_n(\varphi_{n+1}-\varphi_n)
=E z_n p_n (\chi_n-z_n\varphi_n)\,,
\lab{BA-com}
\ee
where $\varphi_n$ and $\chi_n$ are single-valued function of $Q(E)$
on the Riemann surface $\Gamma_N$. We notice that for $E=0$ the
solutions to this equation do not depend on the reggeon number $n$.
Moreover, using \re{L_k} and \re{A} one can show that $L_k(E=0)=\II$,
$A_k^{(\alpha)}(E=0)=0$ %for $3 \le\alpha\le N$
and, therefore, there
exist two special points
on the Riemann surface $Q_0^\pm=(\pm,E=0)$, at which the Baker-Akhiezer
function takes constant values, $\Psi_k(Q_0^\pm;\{t\})={\rm const}$.
The choice of the constants implies the normalization of the Baker-Akhiezer
function.

Let us show that the values $\Psi_k(Q_0^\pm)$ are fixed by the
constraints \re{con}. At the vicinity of $E=0$ on one the sheets
of $\Gamma_N$ the Baker-Akhiezer function can be expanded as
$$
\Psi_n(Q;\{t\})=\Psi^{(0)} + E\,\Psi_n^{(1)}(\{t\}) + \CO(E^2)\,,
$$
where coefficients depend on the choice of sheet and the leading
term $\Psi^{(0)}$ is a constant. Let us substitute this identity into
relation $\Psi_{1+N}=T(E) \Psi_1=\exp(P(E))\Psi_1$ and expand
its both sides in powers of $E$. Taking into
account small-$E$ expansion of the monodromy matrix \re{T},
\be
T(E) = \II + iE \lr{\begin{array}{cr} L_3 & L_- \\ L_+ & -L_3
\end{array}} +\CO(E^2) \,,
\lab{T0}
\ee
one obtains that $\Psi^{(0)}$ is an eigenvector of the following
matrix
$$
\lr{\begin{array}{rc} -h & L_- \\ 0 & h
\end{array}}\Psi^{(0)}=-i P'(0) \Psi^{(0)}\,,
$$
where conditions \re{con} were imposed. We find two
solutions for $\Psi^{(0)}$ corresponding to $P'(0)=\pm ih$, which fix
(up to overall constant) the value of the Baker-Akhiezer functions at the
points $Q_0^\pm$ as
\be
\Psi_n(Q_0^+)=\lr{1 \atop 0}\,,\qquad
\Psi_n(Q_0^-)=\lr{ \frac{L_-}{2h} \atop 1}\,.
\lab{nonn}
\ee
Here, $h$ is the conformal weight of holomorphic wave function and
$L_-=-i\sum_{k=1}^N p_k$.
In the quantum case, the operator $L_-$ commutes with
the hamiltonians $q_k$, but it does not take any definite value on the
subspace \re{con}, since $[L_-,L_3]=2iL_-$ and $[L_-,L_+]=iL_3$.
In the quasiclassical limit, $L_-$ has a zero Poisson bracket with
$q_k$ and it does not depend on the evolution times $t_k$. Its value
is fixed by the initial conditions as
$$
L_-=-i\sum_{k=1}^N p_k(0)\equiv -i P_{\rm tot}\,,
$$
with $P_{\rm tot}$ being the total holomorphic momentum
of $N$ reggeons.%
\footnote{In hadronic scattering amplitudes, $P_{\rm tot}$ is defined
as a holomorphic component of the total momentum transferred in the
$t-$channel.}

We notice,
however, that the relations \re{BA-def} and \re{BA-com} are invariant under
$SL_2$ transformations
$$
p_k\to p_k (c z_k+d)^2\,,\qquad
\Psi_n \to \matrix{d & c \\ b & a} \Psi_n
$$
with the coordinates $z_k$ transformed as in \re{sl2}. Since this
transformation changes the value of $L_-$, one can choose its
parameters in such a way that one puts $L_-=0$ and keeps the
relations $L_3=-h$ and $L_+=0$ unchanged.%
\footnote{These three conditions form a set of the second-class
constraints for the $N$ reggeon system. Their quantization leads to
the Baxter equation for the wave function of the $N$ reggeon compound
state \ci{F}.}
The relation between original and
transformed reggeon coordinates looks as follows
\be
\frac1{z_n}=-i\frac{P_{\rm tot}}{2h}+\frac1{z_n^{_{(0)}}}\,,
\qquad
p_n=p_n^{_{(0)}}\lr{1-i\frac{P_{\rm tot}}{2h}z_n^{_{(0)}}}^2\,.
\lab{tot}
\ee
It allows to integrate the evolution equations \re{EE} for $L_-=0$
to get expressions for $z_n^{_{(0)}}$ and $p_n^{_{(0)}}$
and then restore the physical solutions using \re{tot}.

Therefore, in what follows we put $L_-=0$ and obtain the normalization
conditions \re{nonn} for the Baker-Akhiezer function in the canonical form
(up to overall constant factor)
\be
\Psi_n^{_{(0)}}(Q_0^+)=\lr{1 \atop 0}\,,\qquad
\Psi_n^{_{(0)}}(Q_0^-)=\lr{0 \atop 1}\,.
\lab{BA-nor}
\ee
It easy to see that for $L_-=L_+=0$ the solutions of the evolution
equations \re{EE} and \re{con} are invariant under rescaling of the
coordinates
\be
z_n^{_{(0)}} \to \rho\, z_n^{_{(0)}} \,,\qquad
p_n^{_{(0)}} \to \rho^{-1}\, p_n^{_{(0)}}\,.
\lab{resc}
\ee
For the Baker-Akhiezer function, this corresponds to a freedom in
choosing an overall normalization factors in \re{BA-nor}. One can
further simplify analysis of the evolution equations \re{EE}
for $z_n^{_{(0)}}$ and $p_n^{_{(0)}}$ by noticing that
$q_2=-L_3 L_3=-h^2$ and the dependence
on the ``lowest'' time $t_2$ is given by $\partial_{t_2} z_n^{_{(0)}} =
-2 L_3 \frac{\partial L_3}{\partial p_n} = 2ih z_n^{_{(0)}}$,
and similarly for $p_n^{_{(0)}}$, leading to
\be
z_n^{_{(0)}}(t_2,t_3,...,t_N)=z_n^{_{(0)}}(0,t_3,...,t_N)
\e^{2ih\, t_2}\,,
\qquad
p_n^{_{(0)}}(t_2,t_3,...,t_N)=p_n^{_{(0)}}(0,t_3,...,t_N)
\e^{-2ih\, t_2}\,.
\lab{t2}
\ee
Thus, one can forget for a moment about $t_2-$dependence of the reggeon
coordinates and restore it in the final expressions using \re{t2}.

\subsection{Singularities of the Baker-Akhiezer function}

Let us show that the components of the Baker-Akhiezer function
$\Psi_n(Q)$ are meromorphic functions on the Riemann surface $\Gamma_N$
having $g+1=N-1$ poles, the same number of zeros and essential
singularities at two infinities $Q_\infty^\pm=(\pm,\infty)$ situated
on the upper and lower sheets of $\Gamma_N$. Let us take
the $B-$component of the monodromy matrix \re{T} and observe that
$B(E)/E$ is a polynomial of degree $N-1$ in the spectral parameter $E$
and $B(E) = iL_-\, E + \CO(E^2)$ as $E\to 0$ according to \re{T0}. Then,
one defines the points $E_1$, $...$, $E_{N-1}$ as roots of this
polynomial
$$
B(E_k)/E_k=0\,,\qquad k=1,...,N-1
$$
and considers the relation $T(E)\Psi_1=\exp(P(E))\Psi_1$ at
the points $Q_k=(\pm,E_k)$ on the Riemann surface situated above
$E_k$
\be
\matrix{A(E_k) & 0 \\ C(E_k) & D(E_k)}
\lr{\varphi_1(Q_k) \atop \chi_1(Q_k)}
= \e^{P(E_k)} \lr{\varphi_1(Q_k) \atop \chi_1(Q_k)}\,.
\lab{Tk}
\ee
Solving it one obtains the values of $P(E_k)$
on two sheets of $\Gamma_N$. If the point $Q(E_k)$ belongs to
the same sheet of $\Gamma_N$ on which $D(E_k)=\e^{P(E_k)}$, then one
gets from \re{Tk} that $\varphi_1(Q_k)=0$. Thus, the upper component of
the Baker-Akhiezer function $\Psi_1(Q)$ has $g+1$ zeros on $\Gamma_N$
above the points $E_k$ defined as roots of the polynomial $B(E)/E$.
In similar way, $g+1$ zeros of the lower component of $\Psi_1(Q)$ are
related to the zeros of the polynomial $C(E)/E$ defined by the
$C-$component of the monodromy matrix. In general, two components
of the Baker-Akhiezer function, $\Psi_n^\alpha(Q;\{t\})$ $(\alpha=1,2)$,
have different sets of $g+1$ zeros on $\Gamma_N$ and their positions on
$\Gamma_N$ depend both on the reggeon number, $n$, and the evolution
times, $\{t\}$. Relation \re{BA-nor} implies that one of the roots should
be at the points $Q_0^\pm$. Indeed, it follows from \re{T0} and \re{T}
that the polynomials $B(E)/E$ and $C(E)/E$ vanish at $E=0$ for $L_-=0$ and
$L_+=0$, respectively, and therefore the components of the
Baker-Akhiezer function vanish at the points $Q_0^\pm=(\pm,0)$ in
agreement with \re{BA-com}.

The remarkable property of the roots $E_k=E_k(z_1^{(0)},p_1^{(0)},...,
z_N^{(0)},p_N^{(0)})$ and the corresponding values of $P(E_k)$ is that
they form the set of separated coordinates for the system of $N$ reggeons
\ci{SoV}. Namely, they belong to the spectral curve
\be
\e^{P(E_k)}+\e^{-P(E_k)}=\Lambda(E_k)=2+ q_2 E_k^2 + ... + q_N E_k^N\,,
\qquad
k=1,...,N-1
\lab{sep}
\ee
and have the following Poisson brackets
$$
\{E_k,E_n\}=\{P(E_k),P(E_n)\}=0 \,,\qquad
\{E_k^{-1},P(E_n)\}=\delta_{kn}\,.
$$
Inverting the relations \re{sep} one can express $q_k$ in terms
of $E_k$ and then obtain the equations of motion for zeros $E_k$ of the
Baker-Akhiezer function generated by hamiltonians $q_k$
in the following differential form \ci{pre}
\be
d t_k=-\sum_{j=1}^{N-1}
\frac{d E_j \, E_j^{k-2}}{\sqrt{\Lambda^2(E_j)-4}}\,,
\qquad k=2,3,...,N\,.
\lab{time}
\ee
These equations can be integrated by the Abel map and their solution
describe linear reggeon trajectories on the Jacobian variety of the
Riemann surface $\Gamma_N$ \ci{pre}. As we will show in Sect.~3, using the
Baker-Akhiezer function one can perform the inverse Abel transformation and
construct the explicit expressions for the reggeon trajectories on $2-$dim
plane of transverse coordinates in terms of Riemann theta-functions.

For the Baker-Akhiezer function to be a well defined meromorphic
function on the Riemann surface the number of its zeros, $g+1=N-1$,
should match the number of simple poles, which we choose to be at the
points $\gamma_1$, $...$, $\gamma_{g+1}$ in general position on
$\Gamma_N$. Moreover, considering the relations \re{BA-def} and \re{BA-com} at
the points $Q=(\pm,E)$ close to $\gamma_k$ on $\Gamma_N$ one finds
that the components of $\Psi_n^{\alpha}(Q;\{t\})$ share the common
set of poles and the positions of $\gamma_1$, $...$, $\gamma_{g+1}$
on $\Gamma_N$ do not depend neither on $n$, nor on times $t$. This
set can be considered as part of the initial data for the evolution
equations \re{EE}.

Finally, let us examine the relation \re{BA-com} in the neighborhood of
punctures $Q_\infty^\pm=(\pm,\infty)$ on two sheets of $\Gamma_N$.
Taking the limit $E\to\infty$ one obtains from \re{BA-com} that the solutions
have the following asymptotics on, say, $(-)$ sheet,
$\varphi_n(Q\to Q_\infty^-)=\CO(E^n)$ and
$\chi_n(Q\to Q_\infty^-)=\CO(E^n)$. Let us require that the solutions to
\re{BA-com} should have a similar behaviour on the upper sheet,
$\varphi_n(Q\to Q_\infty^+)=\CO(E^{-n})$ and
$\chi_n(Q\to Q_\infty^+)=\CO(E^{-n})$. This behaviour is consistent
with \re{BA-com} provided that the r.h.s.\ of \re{BA-com} scales as
$E^0$ at large
$E$, or equivalently $\chi_n(Q_\infty^+)- z_n\varphi_n(Q_\infty^+)=0$
leading to the expression for the reggeon coordinates as a ratio of the
components of the Baker-Akhiezer function at the point $Q_\infty^+$.
Thus, having constructed the Baker-Akhiezer function satisfying
desired asymptotic behaviour one will be able to determine the reggeon
coordinates.

\section{Construction of the Baker-Akhiezer function}

The existence of the Baker-Akhiezer function with the properties
established in the previous section can be deduced from the following
well-known fact in the theory of finite-gap soliton solutions \ci{Kr}.
For a smooth hyperelliptic algebraic curve $\Gamma_N$ of genus $g=N-2$
with 2 punctures at $Q_\infty^\pm=(\pm,\infty)$ and a given set of
$g+1$ points $(\gamma_1,\gamma_2,...,\gamma_{g+1})$ in general
position there exists a unique function $\Psi^\alpha_n(Q;\{\tau\})$
such that
\begin{description}
\item[--]$\Psi_n$ is a meromorphic outside the punctures $Q_\infty^\pm$
     and has simple poles at the points $\gamma_1,...,\gamma_{g+1}$;
\item[--]$\Psi_n$ satisfies the normalization conditions \re{BA-com}
     at the points $Q_0^\pm=(\pm,E=0)$;
\item[--]At the vicinity of two punctures on the upper and lower sheets
     the components $\Psi^\alpha_n$ $(\alpha=1,2)$
     have the following asymptotics
     \be
     \Psi^\alpha_n(Q\to Q_\infty^\pm;\{\tau\})=E^{\mp n}
     \e^{\pm\sum_{j=1}^{N-2} \tau_j\, E^j}
     \left[
     \phi_\alpha^\pm(n,\{\tau\}) + \CO\lr{\frac1{E}}
     \right]\,,
     \lab{BA-as}
     \ee
     where $\phi_\alpha^\pm$
     is some $E-$independent function.
\end{description}
The following comments are in order. In this definition, the
Baker-Akhiezer function depends on the set of parameters
$\{\tau\}=\tau_1,...,\tau_{N-2}$, which are different, in general, from the
evolution times $\{t\}$ entering \re{EE}. However, as a function of
$\{\tau\}$, $\Psi^\alpha_n(Q;\{\tau\})$ satisfies the system of first-order
linear matrix equations similar to \re{BA-def} and, as we will show
in Sect.~4.1, the evolution times $\{t\}$ corresponding to the hamiltonian
flows are related to auxiliary parameters $\{\tau\}$ by a {\it linear\/}
transformation, $t_{j+2}=\sum_k A_{jk} \tau_k$.

Finally, substituting expression \re{BA-as} into \re{BA-com} and
comparing the
asymptotic behaviour of different sides of the relation \re{BA-com}
for $Q\to Q_\infty^\pm$ we obtain the following consistency conditions
\be
z_n^{_{(0)}}=\frac{\phi_2^+(n,\{\tau\})}{\phi_1^+(n,\{\tau\})}
   =\frac{\phi_2^-(n+1,\{\tau\})}{\phi_1^-(n+1,\{\tau\})}
\lab{soln1}
\ee
and
\be
p_n^{_{(0)}}
=\frac1{z_{n,n+1}^{_{(0)}}}
\frac{\phi_1^+(n,\{\tau\})}{\phi_1^+(n+1,\{\tau\})}
=\frac1{z_{n-1,n}^{_{(0)}}}
\frac{\phi_1^-(n+1,\{\tau\})}{\phi_1^-(n,\{\tau\})}\,.
\lab{soln2}
\ee
Together with \re{tot} these relations provide the solution to the hierarchy
of the evolution equations for reggeon coordinates and momenta, \re{EE}, in
terms of the Baker-Akhiezer function.

\subsection{Basis of differentials}

To write down the expression for the Baker-Akhiezer function satisfying
the above conditions one defines the canonical basis of cycles
on $\Gamma_N$ as shown in Fig.~1 and constructs the basis of normalized
differentials as follows \ci{D}.

The unique set of holomorphic differentials on $\Gamma_N$, or
differentials of the 1st kind, $d\omega_k$  $(k=1,...,N-2)$, is defined as
\be
d\omega_k=\sum_{j=1}^{N-2} \W_{kj} \frac{dE\,E^{j-1}}{y(E)}
         =\sum_{j=1}^{N-2} \W_{kj} \frac{dE\,E^j}{\sqrt{\Lambda^2(E)-4}}\,.
\lab{1st}
\ee
The coefficients $\W_{kj}$ are fixed by the normalization conditions
\be
\oint_{\alpha_j} d \omega_k = 2\pi \delta_{jk}\,,
\qquad j,k=1,...,N-2
\lab{residue}
\ee
and they can be calculated as an inverse to the following matrix
\be
\lr{\W^{-1}}_{jk}=\frac1{2\pi}\oint_{\alpha_k}
\frac{dE\,E^j}{\sqrt{\Lambda^2(E)-4}}\,.
\lab{1/C}
\ee
The unique set of meromorphic differentials on $\Gamma_N$ of the 2nd
kind, $d\Omega^{\j}$ $(j\ge 1)$, with the $j-$th order
pole at the punctures $Q_\infty^\pm$ is defined as
$$
d\Omega^{\j}=\frac{dE}{\sqrt{\Lambda^2(E)-4}}
\left[j q_N E^{N+j-1} + j q_{N-1} E^{N+j-2} + ... + \CO(E) \right]\,.
$$
The coefficients in front of the remaining powers of $E$ in the
numerator are fixed by the normalization conditions
\be
\oint_{\alpha_k} d \Omega^{\j} = 0\,,\qquad
d\Omega^{\j}\bigg|_{Q\to Q_\infty^\pm}\stackrel{E\to\infty}{=}
\pm\lr{jE^{j-1} + \CO(1/E^2)} dE\,.
\lab{2nd-as}
\ee
The unique dipole differentials on $\Gamma_N$, or differentials
of the 3rd kind, $d\Omega_\infty$ and $d\Omega_0$, having simple poles
at the points $Q_\infty^\pm$ and $Q_0^\pm$, respectively, and normalized
by the conditions
\be
\oint_{\alpha_j} d \Omega_\infty = \oint_{\alpha_j} d \Omega_0 =0\,,\qquad
d \Omega_\infty\bigg|_{Q\to
Q_\infty^\pm}\stackrel{E\to\infty}{=}\mp\frac{dE}{E}
\,,\qquad
d \Omega_0\bigg|_{Q\to Q_0^\pm}\stackrel{E\to 0}{=}\pm\frac{dE}{E}
\lab{3rd-as}
\ee
are defined as
\be
d\Omega_\infty=-\frac1{N}\frac{dE\,\Lambda'(E)}{\sqrt{\Lambda^2(E)-4}}
         =\frac{dE}{\sqrt{\Lambda^2(E)-4}}\left[
           -q_N E^{N-1} + ... + \CO(E)\right]
\lab{Om00}
\ee
and
\be
d\Omega_0=\frac{2ihdE}{\sqrt{\Lambda^2(E)-4}}
+ i\sum_{j=1}^{N-2} (\W_0)_j \frac{dE\,E^j}{\sqrt{\Lambda^2(E)-4}}\,.
\lab{Om0}
\ee
To satisfy the normalization conditions \re{3rd-as} the coefficients
$(\W_0)_j$ have to be equal to
\be
(\W_0)_k=-\frac{h}{\pi}\sum_{j=1}^{N-2} \W_{jk} \oint_{\alpha_j}
\frac{dE}{\sqrt{\Lambda^2(E)-4}}
\lab{B}
\ee
with the matrix $\W_{jk}$ defined in \re{1st} and \re{1/C}.

Let us also define the following $(N-2)-$dimensional vectors $V$, $\U^{\j}$
and $A(Q)$
\be
V_k=-i \oint_{\beta_k} d\Omega_\infty\,,\qquad
\U^{\j}_k=-i \oint_{\beta_k} d\Omega^{\j}\,,\qquad
A_k(Q)=\int_{\gamma_0}^Q d \omega_k\,,
\lab{A(Q)}
\ee
where $k=1,...,N-2$, $\gamma_0$ is an arbitrary reference point on
$\Gamma_N$ and integration in $A(Q)$ goes along some path on $\Gamma_N$
between the points $Q$ and $\gamma_0$.

\subsection{Baker-Akhiezer function}

The components of the Baker-Akhiezer function, $\Psi^\alpha_n(Q;\{\tau\})$
$(\alpha=1,2)$, satisfying the conditions formulated in
the beginning of Sect.~3 can be expressed in terms of the
theta-function defined by the Riemann surface $\Gamma_N$
(for definition see Appendix A) as follows \ci{KBBT}
\be
\Psi^\alpha_n(Q;\{\tau\})=\frac{h_\alpha(Q)}{h_\alpha(\widetilde Q_\alpha)}
\frac{\theta(A(Q)+V n + \sum_{j=1}^{N-2} \U^{\j} \tau_j + Z_\alpha)\theta(Z_0)}
     {\theta(A(Q)+Z_\alpha)\theta(V n + \sum_{j=1}^{N-2} \U^{\j} \tau_j + Z_0)}
\exp\lr{n\int_{\widetilde Q_\alpha}^Q d\Omega_\infty +
\sum_{j=1}^{N-2} \tau_j \int_{\widetilde Q_\alpha}^Q d\Omega^{\j}
} \,.
\lab{BA-fin}
\ee
Here, $\widetilde Q_1=Q_0^+$ and $\widetilde Q_2=Q_0^-$ are normalization points,
the $(N-2)-$dimensional constant vectors $Z_\alpha$ and $Z_0$ are given by
$$
Z_\alpha=Z_0-A(\widetilde Q_\alpha)\,,\qquad
Z_0=A(\widetilde Q_1)+A(\widetilde Q_2)-\sum_{n=1}^{g+1}A(\gamma_n)-\CK
$$
with $\CK$ being the vector of Riemann constants. The function
$h_\alpha(Q)$ is defined as
$$
h_\alpha(Q)=\frac{\theta(A(Q)+Z_\alpha)\theta(A(Q)-T_\alpha)}
                 {\theta(A(Q)-T_+)\theta(A(Q)-T_-)}\,,
$$
together with the constants
$$
T_+=\sum_{s=1}^{g-1} A(\gamma_s) + A(\gamma_g) + \CK\,,\quad
T_-=\sum_{s=1}^{g-1} A(\gamma_s) + A(\gamma_{g+1}) + \CK\,,\quad
T_\alpha=\sum_{j=1,2 \atop j\neq \alpha} A(\widetilde Q_j) +
\sum_{s=1}^{g-1} A(\gamma_s) + \CK\,.
$$
The following comments are in order. According to \re{A(Q)},
the definition of the vectors $A(\widetilde Q_\alpha)$ and $A(\gamma_n)$
implies a certain choice of the integration contours connecting the points
$\widetilde Q_\alpha$ and $\gamma_n$ with the reference point
$\gamma_0$. The same contours should enter into the definition of the
contour integrals of meromorphic differentials in the exponent of \re{BA-fin},
$\int_{\widetilde Q_\alpha}^Q d\Omega^{\j}=\int_{\gamma_0}^Q d\Omega^{\j}
-\int^{\widetilde Q_\alpha}_{\gamma_0} d\Omega^{\j}$. Only
under this condition the Baker-Akhiezer function \re{BA-fin} does not
depend neither on the choice of the integration contours, nor on the
reference point $\gamma_0$. Indeed, deforming the integration
contour between, say, points $Q$ and $\widetilde Q_\alpha$ as
$\gamma_{Q\widetilde Q_\alpha}\to\gamma_{Q\widetilde Q_\alpha} +
\sum_j N_j \alpha_j + \sum_j M_j \beta_j$, and using the
transformation properties \re{tran} of the Riemann theta-function one gets that
the variations of the exponent and the prefactor in \re{BA-fin} compensate
each other, while the function $h_\alpha(Q)$ stays invariant.

Let us show that thus defined Baker-Akhiezer function \re{BA-fin}
satisfies necessary normalization conditions.
Considering the expression for $h_1(Q)$ and using the properties of zeros
of the $\theta-$functions, \re{zero}, we notice that two $\theta-$functions
in the denominator of $h_1(Q)$ vanish at the points
$\gamma_1,...,\gamma_{g-1},\gamma_g$
and
$\gamma_1,...,\gamma_{g-1},\gamma_{g+1}$,
respectively, while the numerator vanishes at the points
$\gamma_1,...,\gamma_{g-1},\widetilde Q_2$ plus additional $g$
points coming from the first $\theta-$function. In the expression
for the Baker-Akhiezer function, \re{BA-fin}, the latter cancels,
however, against the same factor in the denominator of \re{BA-fin}
and its zeros are replaced
by zeros of the first $\theta-$function in the numerator of \re{BA-fin}.
Therefore, the component $\Psi^1_n(Q;\{\tau\})$ has simple poles at
the $g+1$ points $\gamma_1, ..., \gamma_{g+1}$ and it vanishes at the
point $\widetilde Q_2\equiv Q_0^-$ plus additional $g$ points $Q_1, ..., Q_g$.
Being solutions of the equation
$\theta(A(Q_s) + Vn + \sum_{j=1}^{N-2} \U^{\j} \tau_j + Z_1)=0$
they satisfy the relation \re{zero}, or
\be
\sum_{s=1}^{N-2} A(Q_s) + A(Q_0^-) - \sum_{s=1}^{N-1} A(\gamma_s)
= - Vn - \sum_{j=1}^{N-2} \U^{\j} \tau_j\,.
\lab{tau}
\ee
One checks that for $Q\to \widetilde Q_1\equiv Q_0^+$ different factors in
\re{BA-fin} cancel against each other leaving us with $\Psi^1_n(Q_0^+)=1$.
As to asymptotic behaviour of $\Psi^1_n(Q)$ for $Q\to Q_\infty^\pm$, it
is controlled by the exponent in \re{BA-fin} and it is in agreement with
\re{BA-as} due to normalization conditions \re{2nd-as} and \re{3rd-as}
for the meromorphic
differentials. The generalization of this analysis to the component
$\Psi^2_n(Q)$ is straightforward. We conclude that \re{BA-fin} gives a unique
expression for the Baker-Akhiezer function corresponding to the $N$
reggeon system.

\section{Explicit solutions}

We examine the asymptotics of the Baker-Akhiezer function \re{BA-fin}
in the vicinity of infinity $Q_\infty^+$ and compare it with \re{BA-as}
to find the explicit expression for the function $\phi^+_\alpha$.
Substituting it into \re{soln1} and \re{soln2}
we obtain the solution to the hierarchy
of the evolution equations for the reggeon coordinates and momenta
in the following form
\be
z_n^{_{(0)}}=\rho \frac
{\theta\lr{V n + \sum_{j=1}^{N-2} \U^{\j}\tau_j + Z_+ + i\Delta}}
         {\theta\lr{V n + \sum_{j=1}^{N-2} \U^{\j}\tau_j + Z_+}}
\exp\lr{i V_0\, n + i \sum_{j=1}^{N-2}  \U_0^{\j} \tau_j}
\lab{z0}
\ee
and
\be
p_n^{_{(0)}}z_{n,n+1}^{_{(0)}}=q_N^{1/N}
\frac{\theta\lr{V n + \sum_{j=1}^{N-2} \U^{\j}\tau_j + Z_+}}
     {\theta\lr{V n + \sum_{j=1}^{N-2} \U^{\j}\tau_j + Z_0}}
\frac{\theta\lr{V (n+1) + \sum_{j=1}^{N-2} \U^{\j}\tau_j + Z_0}}
     {\theta\lr{V (n+1) + \sum_{j=1}^{N-2} \U^{\j}\tau_j + Z_+}}\,.
\lab{p0}
\ee
Here, the constant $\rho$ is given by a ratio of $\theta-$functions
and its value can be made arbitrary using the symmetry \re{resc}.
The $(N-2)-$dimensional constant vector $(Z_+)_k$ is equal to
$$
Z_+=A(Q_\infty^+)+Z_1=A(Q_\infty^+)-A(Q_0^+)+Z_0=\int_{Q_0^+}^{Q_\infty^+}
d\omega + Z_0\,.
$$
The phase shift $\Delta$ and the oscillator frequencies are defined as
\be
\Delta_k=-i\int_{Q_0^-}^{Q_0^+} d \omega_k\,,\qquad
V_0=-i\int_{Q_0^-}^{Q_0^+} d \Omega_\infty\,,\qquad
\U_0^{\j}=-i\int_{Q_0^-}^{Q_0^+} d \Omega^{\j}\,,
\lab{W}
\ee
where the normalized differentials are integrated along the path on
$\Gamma_N$ connecting the points $Q_0^\pm$ and going through the
reference point $\gamma_0$. Although $\gamma_0$ enters into the definition
\re{A(Q)} of the vector $A(Q)$, neither the Baker-Akhiezer function
\re{BA-fin}, nor the reggeon coordinates \re{z0} and \re{p0} depend on its
particular choice. Prefactor, $q_N^{1/N}$, in \re{p0}
originates from the following expansion
$$
\exp\lr{-\int_{Q_0^+}^Qd\Omega_\infty} \stackrel{Q\to Q_\infty^+}{=}
\exp\lr{\frac1{N}\int_0^E \frac{d\Lambda(E)}{\sqrt{\Lambda^2-4}}}
=E q_N^{1/N} (1+ \CO(1/E))
$$
and it is consistent with the definition \re{q_k} of the hamiltonian
$q_N$, that is $\prod_{k=1}^N p_n^{_{(0)}}z_{n,n+1}^{_{(0)}} = q_N$.
The expressions \re{z0} and \re{p0} involve $N-1$ free parameters --
prefactor $\rho$ and $(N-2)-$dimensional vector $Z_+$ (or equivalently
$Z_0$).

Let us show that solutions \re{z0} and \re{p0} satisfy the periodicity
conditions \re{per}. Using transformation properties of the
$\theta-$function, \re{tran}, the same conditions can be expressed as
$$
V_k=-i\oint_{\beta_k}d\Omega_\infty=2\pi \frac{m_k}{N}\,,
\qquad
V_0=-i\int_{Q_0^-}^{Q_0^+}d\Omega_\infty=2\pi \frac{m}{N}\,,
$$
with $m_k$ and $m$ being integers. To verify them we take
into account the relation between eigenvalues of the
monodromy matrix, \re{det}, and the dipole differential \re{Om00}
\be
d P(E) = - N d \Omega_\infty\,,
\lab{PE}
\ee
and calculate from \re{det} and \re{bran} the multivalued function
$P(E)$ at the branching points $e_j$ of the curve $\Gamma_N$ as
$\e^{P(e_k)}=\pm 1$ and at the points $Q_0^\pm$ above $E=0$ as
$\e^{P(0)}=1$.

\subsection{Evolution times}

We recall that parameters $\tau_k$ entering \re{z0} and \re{p0}
do not coincide with the evolution times $t_k$ in \re{EE}.
To find the relation between them we notice that
$g+1$ points $Q_1$, $...$, $Q_g$, $Q_{g+1}\equiv Q_0^-$ obey
\re{tau} being zeros of the component $\Psi^1_n(Q)$ of the Baker-Akhiezer
function. Moreover, the same points correspond to zeros of the
$B-$component of the monodromy matrix and their $t-$evolution is
described by \re{time}. Comparing the relations \re{tau} and \re{time} and
using the definition \re{A(Q)} of the vector $A(Q_s)$ together with
\re{1st} one obtains that up to a unessential additive constant
\be
\sum_{j=1}^{N-2} \U^{\j}_k \tau_j = \sum_{j=1}^{N-2} \W_{kj} t_{j+2}\,,
\lab{tt}
\ee
where matrix $\W$ was defined in \re{1st} and \re{1/C}. Let us also consider
the $\tau-$dependent part of the exponent in \re{z0} and take into account
the relations \re{id} and \re{tt} to get
$$
i\sum_{j=1}^{N-2} \U_0^{\j}\tau_j
%\int_{Q_0^-}^{Q_0^+}d\Omega^{\j} \tau_j
=-\frac{ih}{\pi}\sum_{j,k=1}^{N-2}
\U^{\j}_k \tau_j \oint_{\alpha_k}\frac{dE}{\sqrt{\Lambda^2(E)-4}}
=-\frac{ih}{\pi}\sum_{j,k=1}^{N-2} \W_{kj}
\oint_{\alpha_k}\frac{dE}{\sqrt{\Lambda^2(E)-4}} t_{j+2}\,.
$$
Comparing this identity with \re{B} and \re{W} one obtains
\be
\sum_{j=1}^{N-2} \U_0^{\j} \tau_j = \sum_{j=1}^{N-2} (\W_0)_j t_{j+2}\,.
\lab{tt1}
\ee
Substituting transition formulas \re{tt} and \re{tt1} into \re{z0} and
\re{BA-fin} we restore the dependence of the reggeon coordinates and the
Baker-Akhiezer function on the evolution times $t_k$. Finally,
combining together \re{z0}, \re{tt}, \re{tt1} and \re{t2} we obtain the
expressions for the holomorphic reggeon coordinates as
\be
z_n^{_{(0)}}(t)= \rho
\exp\lr{2ih\, t_2 +i V_0\, n + i \W_0 \cdot t}
\frac
{\theta\lr{V n + \W\cdot t+ Z_+ + i\Delta}}
{\theta\lr{V n + \W\cdot t + Z_+}}\,,
\lab{soln}
\ee
where the following notations were introduced
$$
\W_0\cdot t = \sum_{j=1}^{N-2} (\W_0)_j t_{j+2}\,,\qquad
\W_k\cdot t=\sum_{j=1}^{N-2} \W_{kj} t_{j+2}\,.
$$
The ratio of $\theta-$functions in \re{soln} describes linear trajectories
on the Jacobian $\CJ(\Gamma_N)$ of the curve \re{curve}.
Relation \re{soln} implies
that the motion of reggeons becomes an image of these trajectories on the
complex $z-$plane modulated by $U(1)$ rotations with the basic frequencies
$(\W_0)_k$.

\subsection{Curves with real branching points}

In the above consideration, the conserved charges $q_k$ as well as
the branching points $e_k$ of the curve $\Gamma_N$ were assumed to
have arbitrary complex values. Let us consider the situation when
all $2N-2$ branching points \re{bran} of the algebraic curve $\Gamma_N$
are distinct and real,
\be
e_1 < ...  < e_{2N-2}\,.
\lab{real}
\ee
The branching points were defined in \re{bran} as roots of the polynomial
$\Lambda^2(E)-4$ and in order for $e_k$ to be real the values of
charges $q_k$ should also be real and, most importantly, they have
to satisfy certain additional conditions. In particular,
$q_2$ should be negative
\be
q_2 < 0\,,
\lab{q2<0}
\ee
while the explicit form of the conditions on $q_3$, $...$, $q_N$ depends
on the number of reggeons $N$ \ci{Qua,pre}. For example, for $N=3$ states,
they can be expressed as
\be
0 \le u_3^2 \le \frac1{27}\,,
\lab{reg3}
\ee
and for $N=4$ states as
\be
-4 u_4 \le u_3^2 \le \frac89
\frac{u_4\lr{\sqrt{48 u_4+1}-2}^2}{\sqrt{48 u_4+1}-1}\,,
\lab{reg4}
\ee
where $u_3=q_3/(-q_2)^{3/2}$ and $u_4 = q_4/q_2^2$.
Let us identify the moduli of the hyperelliptic curve $\Gamma_N$ as
$(N-2)$-dimensional vector with the components $u_k=q_k/(-q_2)^{k/2}$
 $(k=3,...,N)$. The relations \re{reg3} and \re{reg4}
define the regions on the moduli space, $\CM(\Gamma_N)$, corresponding
to curves $\Gamma_N$ with real
branching points. At the boundary of these regions two of the branching
points merge and the curve $\Gamma_N$ becomes singular.

There is a special interest in considering soliton solutions for
curves with real branching points. We recall, that the reggeon
trajectories \re{z0} appear as solutions of the evolution equations
generated by the leading term $S_0$ of the WKB expansion of the
wave function of $N$ reggeon state \re{WKB}. A natural question arises
whether the WKB expansion gives a meaningful approximation to solutions
of the Schr\"odinger equation \re{BKP}. It turns out that for the quantum
numbers $q_k$ satisfying \re{reg3} and \re{reg4} one can construct the
{\it exact\/} solutions to \re{BKP} within the Bethe Ansatz approach
\ci{FK,Bet}. A throughout analysis shows \ci{Qua} that the WKB expansion
is in a good agreement with the exact solutions. Thus, similar to electron
orbits in the Bohr model of hydrogen atom, the reggeon trajectory,
corresponding to curves $\Gamma_N$ with real branching points, describe
the quasiclassical limit of the quantum states constructed in \ci{FK,Bet}.

Let us consider the properties of solitons \re{z0} under additional
condition \re{real}. According to our definition of the canonical set
of cycles shown in Fig.~1, the cycle $\alpha_j$ lies on $\Gamma_N$ above
the interval $[e_{2j},e_{2j+1}]$, called forbidden zones, on which
$\Lambda^2(E)>4$. The Riemann surface defined by the curve $\Gamma_N$ is
constructed by gluing two copies of the complex $E-$plane along $2N-2$
finite intervals $[e_{2j},e_{2j+1}]$ and one infinite interval
$[e_{2N-2},e_1]$ going through infinity. On the intervals $[e_{2j-1},e_{2j}]$
$(j=1,...,N-1)$, called allowed zones, one has $\Lambda^2(E)-4 \le 0$ and
the eigenvalue of the monodromy matrix, $P(E)$, has pure imaginary values
with $\frac1{N}\Im P(E)$ being a quasimomentum. In particular, since
$\Lambda(0)=2$, the point $E=0$ belongs to one of the allowed zones,
which we denote as $[e_{2K-1},e_{2K}]$,
\be
e_1 < ... < e_{2K-1} < 0 < e_{2K} < ... < e_{2N-2}\,.
\lab{K}
\ee
The values of $e_{2K-1}$ and $e_{2K}$ can be defined as roots of
\re{bran} closest to the origin. The points $Q_0^\pm$ belong to the
different sheets of $\Gamma_N$ and one has to specify
the integration path on $\Gamma_N$ between the points $Q_0^+$ and $Q_0^-$
entering \re{W}. We would like to stress that
although its choice in \re{W} can be arbitrary, in order to apply the
relations \re{tt} and \re{tt1} one has to require that the path should not
cross the canonical cycles $\alpha_j$ and $\beta_j$. The later
condition fixes it uniquely as shown in Fig.~1. The path is trapped
between the cycles $\beta_{K-1}$ and $\beta_K$ and its orientation on
$\Gamma_N$ is opposite to that of the $\beta-$cycles.

Under these definitions it is easy to see from \re{1/C},
and \re{B} that parameters $\W_{jk}$, $(\W_0)_k$, $V$, $V_0$ and $\Delta$
are real. Moreover, according to \re{PE} and \re{A(Q)}, the components of
the vector $V_n$ are given by the difference of values of quasimomentum at the
branching points $e_{2N-2}$ and $e_{2n+1}$, that is
$$
V_n=\left\{
\begin{array}{ll}
 -\frac{2\pi}{N}(N-n)\,, & 1 \le n \le K-1 \\[3mm]
 -\frac{2\pi}{N}(N-n-1)\,, & K \le n \le N-2
  \end{array}
 \right.\,,
$$
where integer $K$ is defined by configuration of the branching points.
In a similar way, the parameter $V_0$ can be calculated as
$$
V_0=2i\int_0^{e_{2K}} d\Omega_\infty + i\oint_{\beta_{K}}d\Omega_\infty
   = \frac{2\pi}{N} - V_K = \frac{2\pi}{N}(N-K)\,.
$$
For the vector $\Delta_n$ we get
$$
\Delta_n=2i\int_0^{e_{2K}} d\omega_n+i\oint_{\beta_{K}}d\omega_n
        = \Delta^{_{(0)}}_n + 2\pi i \tau_{Kn}
$$
where $\tau_{Kn}$ is an element of Riemann matrix, \re{R-B}, and
$\Delta^{_{(0)}   }_n=2\sum_{j=1}^{N-2} \W_{nj}\int_0^{e_{2K}}
\frac{dE\,E^j}{\sqrt{4-\Lambda^2(E)}}$.

\subsection{Solitons for $N=3$ reggeon states}

Let us consider in more detail the properties of the soliton waves for
$N=3$ reggeon states with the quantum numbers $q_2=-h^2$ and $q_3$
satisfying the conditions \re{q2<0} and \re{reg3}. Choosing in \re{reg3}
the branch $0 \le q_3/h^3 \le \frac1{\sqrt{27}}$ one matches the branching
points of $\Gamma_3$ into \re{K} to get $K=1$.
The soliton solution \re{soln} takes the form
\be
z_n^{_{(0)}}(t_2,t_3)=\rho \exp\lr{2ih\,t_2+\frac{4i\pi}{3}n+ i \W_0 t_3}
\frac{\theta(-\frac{2\pi}{3}n+\W t_3+Z_++i\Delta)}
{\theta(-\frac{2\pi}{3}n+\W t_3+Z_+)}\,,
\lab{soln-3}
\ee
where $\rho$ and $Z_+$ are arbitrary complex parameters defined
by initial conditions. The values of $\W_0$, $\W$ and $\Delta$ can
be calculated from \re{1/C}, \re{B} and \re{W} as elliptic integrals
\be
\W=\frac{\pi}{\int_{e_2}^{e_3}\frac{dE\,E}{\sqrt{\Lambda^2-4}}}\,,
\qquad
\W_0=-2h\frac{\int_{e_2}^{e_3}\frac{dE}{\sqrt{\Lambda^2-4}}}
         {\int_{e_2}^{e_3}\frac{dE\,E}{\sqrt{\Lambda^2-4}}}\,,
\qquad
\Delta=2 \W\int_{e_1}^0\frac{dE\,E}{\sqrt{4-\Lambda^2}}
\lab{BC}
\ee
with $\Lambda(E)=2-h^2 E^2+q_3 E^3$ and $\Lambda(e_1)=\Lambda(e_2)=
\Lambda(e_3)=-2$. The modular parameter $\tau$ entering the definition
of the Riemann $\theta-$function takes pure imaginary values,
$\Im\tau \ge 0$,
\be
\tau=i\frac{\int_{e_3}^{e_4}\frac{dE\,E}{\sqrt{4-\Lambda^2}}}
           {\int_{e_2}^{e_3}\frac{dE\,E}{\sqrt{\Lambda^2-4}}} \,.
\lab{mod}
\ee
For $q_3$ inside the region \re{reg3}, the expression \re{soln-3}
defines quasiperiodic reggeon trajectories on the $(x,y)-$plane of
transverse coordinates, $z=x+iy$. An example of such trajectory
corresponding to the quantum numbers $q_2=-1$ and $q_3=1/7$ is shown
in Fig.~2.
\begin{figure}[ht]
\vspace*{-15mm}
\centerline{\epsffile{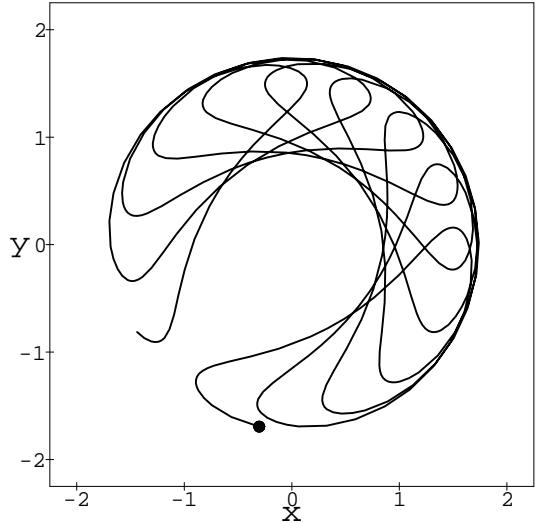}}
\vspace*{-15mm}
\caption{The trajectory of the reggeon on $(x,y)-$plane
described by eq.\re{soln-3} for $t_2=0$ and $n=1$. The parameters
have the following values: $\rho=1$, $h=1$ and $Z_+=3-i$.
Dot denotes the initial position of the reggeon at $t_3=0$.}
\end{figure}
Interesting property of the trajectories \re{soln-3} is that at any moment
of time $t_2$ and $t_3$ the reggeon coordinates satisfy the condition
\be
 \frac{z_1-z_2}{z_1+z_2}
+\frac{z_2-z_3}{z_2+z_3}
+\frac{z_3-z_1}{z_3+z_1}
=
-\frac{(z_1-z_2)(z_2-z_3)(z_3-z_1)}
      {(z_2+z_3)(z_2+z_3)(z_3+z_1)}
=i\frac{h^3}{q_3}\,.
\lab{nice}
\ee
It follows from the expression \re{q_k}, $q_3=z_{12}z_{23}z_{31}p_1p_2p_3$,
after one excludes the momenta using $L_3=-h$ and $L_+=L_-=0$.

For $q_3=0$ and $q_3=h^3/\sqrt{27}$ the branching points of the curve
$\Gamma_3$ merge leading to $\tau=0$ and $\tau=i\infty$, respectively.
The curve $\Gamma_3$ becomes singular and the properties of soliton
solutions are changed.

\subsubsection{Singularity at \mbox{\boldmath $q_3=\frac{h^3}{\sqrt{27}}$}}

In this case, the branching points are located at $e_1=-\sqrt3/h$,
$e_2=e_3=2\sqrt{3}/h$, $e_4=3\sqrt3/h$ and the $\alpha-$cycle on
$\Gamma_3$ is shrinking into a point. Applying relations \re{BC} and
\re{mod} one gets
\be
\W_0=-\W=-\frac{h^2}{\sqrt 3}\,,\qquad \Delta=-2\ln 2\,,\qquad
\tau\to i\infty\,.
\lab{B-C}
\ee
We notice that in the limit $\tau\to i\infty$ and $Z_+=\CO(\tau^0)$
the $\theta-$functions in \re{soln-3} are replaced by $1$, since only
one term with $m=0$ survives in the infinite sum \re{theta}.
The corresponding solution,
$
z_n^{_{(0)}}(t_2,t_3)=\rho\exp\lr{2iht_2+\frac{4\pi i}3 n-i\frac{h^2}{\sqrt
3}t_3}\,,
$
describes three reggeons located in the vertices of equilateral triangle
and rotating around the origin as a whole. In order to avoid such rigid
reggeon configurations, one has to adjust the constant $Z_+$ as
$$
Z_+=\pi \tau + Z_\infty + \CO(\tau^{-1})
$$
and use the asymptotic behaviour of the $\theta-$function \re{theta}
$$
\theta(u+\pi\tau;\tau)=1+\e^{-iu}\,,\qquad
\mbox{as $\tau\to i\infty$}\,.
$$
Finally, the soliton solution for $q_3=h^3/\sqrt{27}$ becomes
\be
z_n^{(0)}(t_2,t_3)=\rho \e^{2iht_2+\frac{2\pi i}3 n}
\frac{4+\e^{i\Phi}}{4(1+\e^{-i\Phi})}\,,
\qquad
\Phi=\frac{2\pi}3n-\frac{h^2}{\sqrt 3}t_3-Z_\infty
\lab{27}
\ee
with $\rho$ and $Z_\infty$ being arbitrary constants. This
relation defines periodic reggeon trajectories on the plane
and example of such trajectory is shown in Fig.~3.

\begin{figure}[ht]
\vspace*{-15mm}
\centerline{\epsffile{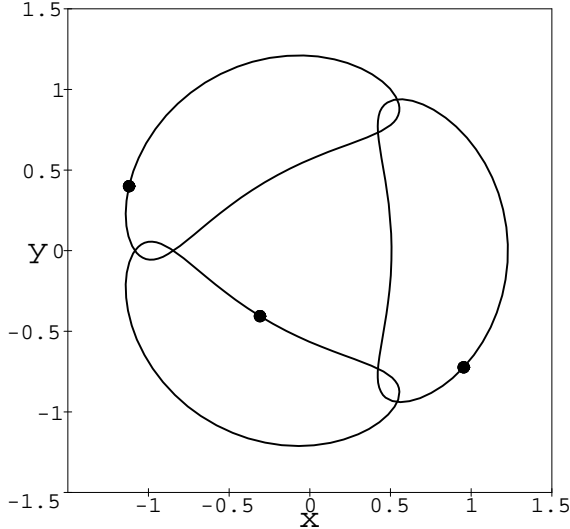}}
\vspace*{-15mm}
\caption{The reggeon trajectories on $(x,y)-$plane described by eq.\re{27}
for $t_2=0$. The parameters have the following values:
$\rho=1$, $h=1$ and $Z_\infty=1+i$. Dots denote initial positions of
reggeons at $t_3=0$.}
\end{figure}

\subsubsection{Singularity at \mbox{\boldmath $q_3=0$}}

In this case, $e_1=-e_2=-2/h$ and two remaining branching points merge
at infinity, $e_3=e_4=\CO(1/q_3)$. Integration in \re{BC} and \re{mod}
yields
$$
\W=-i\tau h^2\,,\qquad
\Delta=i\tau \pi\,,\qquad
\tau\to 0\,.
$$
To analyze \re{soln-3} at small $\tau$ one uses the duality property of the
$\theta-$function \ci{D}
\be
\theta(z;\tau)=\lr{-i\tau}^{-1/2}\exp\lr{-\frac{iz^2}{4\pi\tau}}
\theta\lr{\frac{z}{\tau};-\frac1{\tau}}\,,
\lab{dual}
\ee
which maps $\tau\to 0$ limit into asymptotic behaviour
at $\tau\to -i\infty$. Taking the constant $Z_+$ in the
form $Z_+=Z_0+\tau Z_1$ with $Z_0$ and $Z_1$ arbitrary one gets from \re{dual}
and \re{soln-3} that in the limit $\tau\to 0$ the solution \re{soln-3}
becomes $t_3-$independent $z_n^{_{(0)}}(t_2,t_3)= (-)^n \rho
\exp(2hit_2+\frac{i}2 Z_0)$. To get a nontrivial solution one has to
adjust the constant $Z_+$ as
$$
Z_+=\frac{\pi}3 + \tau Z_1 + \CO(\tau^2)\,.
$$
The resulting expression for reggeon coordinates looks like
\be
z_1^{(0)}(t_2,t_3)=-z_3^{(0)}(t_2,t_3)=\rho \e^{2iht_2}\,,\qquad
z_2^{(0)}(t_2,t_3)=z_3^{(0)}(t_2,0)\tanh\lr{\frac{h^2}2t_3-\frac{i}2Z_1}\,.
\lab{0}
\ee
Here, the positions of reggeons with the numbers $n=1$ and $n=3$ are frozen
in time $t_3$ and the $n=2$nd reggeon moves between them,
$z_2^{(0)}(t_2,-\infty)=z_1^{(0)}(t_2,0)$ and
$z_2^{(0)}(t_2, \infty)=z_3^{(0)}(t_2,0)$. It is easy to see that
the trajectory \re{0} is consistent with \re{nice}.

\section{Multi-soliton solutions}

As we have seen in the previous section, the soliton solutions
corresponding to the degenerate singular spectral curve $\Gamma_N$ can
be written for $N=3$ in terms of trigonometric functions. In the
finite-gap theory they are called multi-soliton solutions \ci{NMPZ}.
Let us show that the same property holds for any $N$. Although
multi-soliton solutions can be obtained as limits of a general periodic
solutions \re{soln} for degenerate
curve $\Gamma_N$, there is a simpler way to write them directly without
going through a complicated degeneration procedure.

The multi-soliton solutions correspond to a degenerate case, when all but
two roots of the equation \re{bran} are double, that is
\be
\Lambda^2(E)-4 = q_N^2\, (E-e_-)(E-e_+) E^2 \prod_{j=1}^{N-2} (E-e_j)^2\,,
\lab{e+-}
\ee
with $e_-\neq e_+$. In this limit, the branching points of the curve $\Gamma_N$
merge and the definition \re{curve} of $\Gamma_N$ takes the form
\be
\Gamma_N^{\rm sing}:\qquad
\widetilde y^2=(E-e_-)(E-e_+)\,,\quad
y = \widetilde y \times q_N\prod_{j=1}^{N-2} (E-e_j)\,.
\lab{ytilde}
\ee
Here, $\widetilde y$ is a rational curve and the Riemann surface corresponding to
$\Gamma_N^{\rm sing}$ has a genus $g=0$, a sphere. It is constructed by gluing
together two sheets along the cuts $(\infty,e_-]$ and $[e_+,\infty)$.
One can introduce a global complex parameter $\k$ on $\Gamma_N$ and
parameterize the solutions to \re{ytilde} as
\be
E=\beta\lr{\k+\frac1{\k}} + \alpha\,,\qquad
\widetilde y=\beta\lr{\k-\frac1{\k}}
\lab{k}
\ee
with $\beta=\frac14(e_+-e_-)$ and $\alpha=\frac12(e_++e_-)$. In this
parameterization, the points $\k=\infty$ and $\k=0$ correspond to the
punctures $Q_\infty^\pm$ on the upper and lower sheets, respectively.
The permutation of sheets of $\Gamma_N$ corresponds to involution in
$\k-$plane, $\k\to 1/\k$, that does change the value of $E$,
\be
E(\k)=E(1/\k)\,.
\lab{1/k}
\ee
All meromorphic functions (and differentials) on $\Gamma_N$ introduced
before can be now rewritten as functions of the parameter $\k$ only. In
particular, consider the function $w=\e^{P(E)}$, the Bloch multiplier,
defined in \re{Bloch}. According to \re{Bl}, it has a pole of order $N$ at
$Q_\infty^+$ and a zero of degree $N$ at $Q_\infty^-$. Therefore, as
a function of $\k$ it has a unique form $w=a \k^N$. At the vicinity of
the punctures $Q_\infty^\pm$ one substitutes $\k\sim \beta/E \to 0$ and
$\k\sim E/\beta \to \infty$ into $w(\k)$ and matches the asymptotics of
$w(E)$ with \re{Bl} to get the values of the constants, $a=1$ and
$\beta=q_N^{-1/N}$, leading to
$$
w=\e^{P(E)}=\k^N\,.
$$
To find positions of the branching points of $\Gamma_N^{\rm sing}$ we
substitute this relation into \re{det},
\be
\Lambda(E)=\k^N + \k^{-N}= 2 + q_2 E^2(\k) + ... + q_N E^N(\k) \,,
\lab{E(k)}
\ee
and notice that, according to \re{e+-}, $\Lambda^2(E)=4$ at the
branching points and $\Lambda(E)=2$ for $E=0$. Thus, $\k^{2N}=1$ at the
branching points $e_\pm$ and $e_j$, while $\k^N=1$ at two points $Q_0^\pm$
on $\Gamma_N^{\rm sing}$ above $E=0$. These conditions define $2N$ points
on the complex $\k-$plane
\be
\k_j=\e^{\frac{i\pi}{N}j}\,,
\qquad
(j=1,...,2N-1)
\lab{kj}
\ee
and we identify the corresponding values
$E(\k_j)=2q_N^{-1/N}\cos\lr{\frac{\pi j}{N}}+\alpha$ as $2N$ roots of
$\Lambda^2-4$ in \re{e+-}. Since $1/\k_j=\k_{2N-j}$, it follows from \re{1/k}
that, in agreement with \re{e+-}, $2N-2$ roots are degenerate,
$E(\k_j)=E(\k_{2N-j})$ and
$j\neq 0\,, N$. The remaining two roots, $j=0\,,N$, correspond
to the branching points $e_+=E(\k_0)$ and $e_-=E(\k_N)$. Among $N-1$ pairs
$\k_j$, $\k_{2N-j}$ of the double roots one has to specify the pair that
corresponds to two points $Q_0^\pm$ on $\Gamma_N^{\rm sing}$ above $E=0$.
Let $j=j_0$ be the corresponding index. Since $e_\pm\neq E(\k_{j_0})=0$
and $\k_{j_0}^N=1$ the values of $j_0$ are constrained to be positive
even
\be
0 < j_0 < N\,, \qquad j_0/2=\IZ\,.
\lab{j0}
\ee
This index parameterizes different multi-soliton solutions. For given
$j_0$ one solves the equation $E(\k_{j_0})=0$ to find the constant
$\alpha=-2q_N^{-1/N}\cos(\pi j_0/N)$ and obtain the branching points of
$\Gamma_N^{\rm sing}$ as
$$
e_j=2q_N^{-1/N}\left[\cos\lr{\frac{\pi j}{N}}-\cos\lr{\frac{\pi j_0}{N}}\right]
\,,\qquad
e_\pm=2q_N^{-1/N}\left[\pm1-\cos\lr{\frac{\pi j_0}{N}}\right]
$$
with $j=1,...,N-1$ and $j\neq j_0$. The branching points fix uniquely
the curve $\Gamma_N^{\rm sing}$ and allow to evaluate the
corresponding values of the quantum numbers $q_k$. Substituting
the first relation \re{k} into \re{E(k)} and comparing the coefficients
in front of different powers of $\k$ in the both sides of \re{E(k)}
one obtains the system of equations on $q_k$ and, in particular,
$$
q_2 q_N^{-\frac2{N}}=-\frac{N^2}{4\sin^2\lr{\frac{\pi j_0}{N}}}\,,
\quad
...
\quad
q_{N-1}q_N^{-\frac{N-1}{N}}=2N\cos\lr{\frac{\pi j_0}{N}}\,.
$$
For $N=3$ and $N=4$ the solution to \re{j0} is $j_0=2$ leading to
$q_2 q_3^{-2/3}=-3$ and $(q_2 q_4^{-1/2}=-4,\,q_3=0)$, respectively.

Let us consider the Baker-Akhiezer function $\Psi_n^\alpha$ on
$\Gamma_N^{\rm sing}$. Its properties formulated in Sect.~3
imply that being functions of complex $\k$ it can be written in
the form
\be
\Psi_n^\alpha(\k;\{\tau\})=\k^{-n} \e^{\sum_{j=1}^{N-2}\tau_j
\lr{\k^j-\k^{-j}}\,q_N^{-j/N}}\times\frac{R_\alpha(\k;n,\{\tau\})}{R_0(\k)}\,,
\lab{BA-MS}
\ee
where $R_0$ and $R_\alpha$ $(\alpha=1,2)$ are polynomials of degree
$N-1$ in $\k$. The zeros of $R_0(\k)$ become poles of $\Psi_n^\alpha$
and in what follows can be considered as parameters of the multi-soliton
solutions,
$
R_0(\k)=\prod_{s=1}^{N-1} (\k-\gamma_s)\,.
$
To verify the asymptotic behaviour \re{BA-as} one has to take into account
the relation
\be
E(\k)=q_N^{-1/N}\lr{\k+\frac1{\k}-2\cos\lr{\frac{\pi j_0}{N}}}
\lab{Ek}
\ee
and perform the limits $\k\to 0$ and $\k\to\infty$. The unknown so far
polynomials $R_\alpha$ are uniquely defined by the normalization
conditions \re{BA-nor},
\ba
\Psi_n^1(1/\k_{j_0})=1\,,\qquad \Psi_n^1(\k_{j_0})=0
\lab{MS-nor}
\\
\Psi_n^2(1/\k_{j_0})=0\,,\qquad \Psi_n^2(\k_{j_0})=1
\nonumber
\ea
and additional $N-2$ linear relations
\be
\Psi_n^\alpha(1/\k_j)=\Psi_n^\alpha(\k_j)\,,
\qquad j=1,...,N-1\,,\quad j\neq j_0,.
\lab{lin}
\ee
The meaning of the last condition is that parameters $\k_j$ and $1/\k_j$
define the same point on the curve. It belongs simultaneously to both
sheets of $\Gamma_N^{\rm sing}$ and the value of the functions
$\Psi_n^\alpha$ should be the same. Let us choose
the polynomial $R_2(\k;n\{\tau\})$ in the following form
$$
R_2(\k;n,\{\tau\})=A\times \lr{\k-1/{\k_{j_0}}}
\left[\prod_{j=1\atop j\neq j_0}^{N-1}(\k-\k_j)+(\k-\k_{j_0})
\sum_{l=0}^{N-3} a_l\, \k^{N-3-l}\right]\,,
$$
with $A$ and $a_l$ $(l=0,...,N-3)$ being some functions of $n$
and $\tau$. One verifies that $\Psi_n^2$ satisfies the normalization
conditions \re{MS-nor} provided that $A$ is given by
\baa
A(n,\{\tau\})&=&
\frac{R_0(\k_{j_0})}{(\k_{j_0}-1/\k_{j_0})\prod_{j=1\atop j\neq j_0}^{N-1}
(\k_{j_0}-\k_j)}
\times
\k_{j_0}^n \e^{-\sum_{s=1}^{N-2}\tau_s
\lr{\k^s_{j_0}-\k^{-s}_{j_0}}\,q_N^{-s/N}}
\\
&=&{\rm const}\times
\exp\left[in\frac{\pi j_0}{N} -2i\sum_{s=1}^{N-2}\tau_s
\sin\lr{\frac{\pi j_0s}{N}}\,q_N^{-s/N}\right]\,.
\eaa
The conditions \re{lin} lead to the linear system of equations for $a_l$
that can be expressed as
\be
\sum_{s=0}^{N-3} a_s
\sin\lr{\frac12\Phi_j +\frac{\pi j}{N}s}
=\frac{i}2 \e^{\frac{i}2\Phi_j}
\frac{\prod_{l=1}^{N-1}\lr{1-\k_j \k_l}}{\lr{1-\k_j\k_{j_0}}^2}\,,
\qquad
j=1,...,N-2\,.
\lab{cons}
\ee
Here, the parameters $\k_j$ are given by \re{kj} and the phases
are defined as
\be
\Phi_j(n,\{\tau\})=n\frac{2\pi j}{N} -4\sum_{s=1}^{N-2}\tau_s
\sin\lr{\frac{\pi j s}{N}}\,q_N^{-s/N} + Z_j
\lab{phase}
\ee
with $Z_j$ being some constants depending on zeros $\gamma_1$, $...$,
$\gamma_{N-1}$. Solution to \re{cons} can be written explicitly as
a ratio of determinants. Comparing \re{BA-as} with \re{BA-MS} in the limit
$\k\to \infty$ we get
\be
\phi_2^+(n,\{\tau\})=A\times (1+a_0)={\rm const}\times
\e^{\frac{i}2\Phi_{j_0}}
\left[1+a_0(\e^{i\Phi_1},
...,\e^{i\Phi_{N-1}})\right]\,,
\lab{phi2}
\ee
where $a_0$ is rational function of all phases except
$\e^{i\Phi_{j_0}}$.
Repeating similar analysis for the polynomial $R_1$ one arrives at
the relation
\be
\phi_1^+(n,\{\tau\})=A^{-1}\times (1+\bar a_0)=
{\rm const}\times
\e^{-\frac{i}2\Phi_{j_0}}
\left[1+\bar a_0(\e^{-i\Phi_1},
...,\e^{-i\Phi_{N-1}})\right]\,,
\lab{phi1}
\ee
with $\bar a_0$ being a function complex conjugated to $a_0$.

To finish description of the multi-soliton solutions one has to
turn to the evolution times $t_k$ using the transition formulas
\re{tt} and \re{tt1}. Let us consider the
meromorphic differential of the second kind entering the
definitions \re{A(Q)} and \re{W}. It follows from \re{Ek} and \re{2nd-as},
that on the degenerate curve $\Gamma_N^{\rm sing}$ the
differential $d\Omega^{\j}$ has a $N-$th order pole at the
points $\k=0$ and $\k=\infty$. Therefore it can be written as
$$
d\Omega^{\j}=q_N^{-j/N}d\lr{\k^j-\k^{-j}}
$$
with the prefactor fixed by the asymptotics \re{2nd-as}. Substituting
this identity into \re{A(Q)} and \re{W} one gets
\be
\U_k^{\j}=4q_N^{-j/N}\sin\lr{\frac{\pi j k}{N}} \,,
\qquad
\U_0^{\j}=4q_N^{-j/N}\sin\lr{\frac{\pi j_0 j}{N}} \,,
\lab{UW}
\ee
where $k=1,...,N-1$ and $k\neq j_0$.
To evaluate the r.h.s.\ of \re{tt} and \re{tt1} one has to examine the
$E-$expansion of the differentials $d\omega_k$ and $d\Omega_0$
on $\Gamma_N^{\rm sing}$ and identify the coefficients
$\W_{kj}$ and $(\W_0)_k$. We observe that due to \re{e+-} and \re{1st} the
differential $d\omega_k$ has $N-2$ additional poles on
$\Gamma_N^{\rm sing}$ located at the points $E=e_j$. The normalization
condition \re{residue} implies that it has a vanishing residue at all
points except of $E=e_k$, where its residue is equal to $i$.
This gives the following system of equations on $\W_{kj}$
\be
\sum_{j=1}^{N-2} \W_{kj} E^j = 2q_N^{1-1/N}\sin\lr{\frac{\pi k}{N}}
\times \prod_{i=1 \atop i\neq j_0}^{N-1} (E-e_i)\,,
\lab{C-sing}
\ee
valid for any $E$. The analysis of the differential
$d\Omega_0$ can be performed in a similar way with the only difference
that, according to \re{3rd-as},
the residue of $d\Omega_0$ vanishes at all points $E=e_j$
$(j\neq j_0)$ and its residue at $E=e_{j_0}=0$ is equal to $1$,
\be
2h + \sum_{j=1}^{N-2} (\W_0)_j E^j=
2q_N^{1-1/N}\sin\lr{\frac{\pi j_0}{N}}
\times\prod_{i=1 \atop i\neq j_0}^{N-1} (E-e_i)\,.
\lab{B-sing}
\ee
The relations \re{C-sing} and \re{B-sing} allow to calculate the
coefficients $\W_{kj}$ and $(\W_0)_k$ in terms of the branching points $e_j$
and, in particular,
$$
(\W_0)_{N-2}=2\sin\lr{\frac{\pi j_0}{N}}q_N^{1-1/N}\,,
\qquad
\W_{k,N-2}=2\sin\lr{\frac{\pi k}{N}}q_N^{1-1/N}\,.
$$
One checks, that for $N=3$, leading to $j_0=2$, these expressions are
in agreement with
\re{B-C}. Observing the relation $\W_{j_0,k}=(\W_0)_k$, we substitute
\re{UW} into the transition formulas \re{tt} and \re{tt1} to
obtain the phases \re{phase} in the following
form
$$
\Phi_k(n,\{t\})=n\frac{2\pi k}{N}-\sum_{j=1}^{N-2}\W_{kj}t_{j+2}+Z_k\,,
\qquad
 1 \le k \le N-1\,,
$$
with the coefficients $\W_{kj}$ calculated from \re{C-sing}. Finally,
substitution of \re{phi2} and \re{phi1} into \re{soln1} yields the
multi-soliton solutions for the reggeon coordinates
\be
z_n^{_{(0)}}(t)=\rho\e^{2ih t_2+i\Phi_{j_0}}
\frac{1+a_0(\e^{i\Phi_1},...,\e^{i\Phi_{N-1}})}
{1+\bar a_0(\e^{-i\Phi_1},...,\e^{-i\Phi_{N-1}})}
\lab{multi-N}
\ee
with $\rho$ being a constant. For $N=3$ this expression coincides
with \re{27}. Solving \re{cons} for arbitrary $N$ one obtains the
multi-soliton solutions \re{multi-N} in the form of rational
functions of $N$ exponentials, $\e^{i\Phi_j}$.

\section{Conclusions}

In the present paper we studied the asymptotic solutions of the
Schr\"odinger equation \re{BKP} for the color-singlet reggeon
compound states in multi-color QCD. In the Regge limit, a nontrivial
QCD dynamics affects only transversal reggeon degrees of freedom and
the $N$ reggeon compound states look like a system of $N$ interacting
particles on the 2-dimensional plane of transverse reggeon coordinates.
This system possesses a large enough family of conserved charges $q_k$
for the Schr\"odinger equation to be completely integrable. We
identified the eigenvalue of the ``lowest'' charge $q_2$ as a parameter
of asymptotic expansion playing the role of the effective Planck
constant.

In the leading order of the asymptotic expansion, quantum
fluctuations are frozen and quantum mechanical motion of reggeons is
restricted to the classical trajectories driven by the ``action'' function
$S_0$. The conserved charges $q_k$ are replaced by classical functions
on the phase space of $N$ reggeons. They have a mutually vanishing Poisson
brackets and generate the hierarchy of evolution equations for reggeon
coordinates. For $N=2$ reggeon states, solution to the evolution
equations describes 2 reggeons rotating around the center-of-mass with
the angular velocity given by the conformal weight $h$. For $N\ge 3$,
integrals of motion $q_3$, $...$, $q_N$ generate new modes of the
classical reggeon motion which we identified as soliton waves
propagating on the chain of $N$ particles with periodic boundary
conditions. Applying the methods of the finite-gap theory we constructed
the explicit form of the reggeon trajectories in terms of Riemann
theta-functions and studied their properties.

The orbits of the classical reggeon motion are parameterized by the
eigenvalues of the conserved charges $q_k$ $(k\ge 3)$.
They define the moduli of the hyperelliptic curve $\Gamma_N$
and appear as parameters of the soliton solutions. In the leading order
of the asymptotic expansion the eigenvalues of $q_k$ could take
arbitrary complex values. Quantization of $q_3$, $...$, $q_N$
appears as a result of imposing the Bohr-Sommerfeld
quantization conditions on the classical orbits of reggeon motion.
It was shown \ci{Qua} that for $N=3$ reggeon compound
states the results of the WKB expansion for eigenvalue of $q_3$ are
in a good agreement with the exact expressions \ci{FK,Bet} obtained
by means of the algebraic Bethe ansatz \ci{QISM} for integer positive $h$.
Solutions to the Bohr-Sommerfeld conditions \ci{Qua} give the quantized
values of $q_k$ as functions of the conformal
weight $h$ and some additional set of integers $\{l\}=(l_1,...,l_{N-2})$
parameterizing different families of curves on the moduli space,
$q_k=q_k(h;\{l\})$ $(k=3,...,N)$.
These curves can be found as solutions of the
Whitham equations \ci{pre} describing the adiabatic perturbation
\ci{Wh} of the reggeon soliton waves.

\section*{Acknowledgments}

We are most grateful to L.D. Faddeev for stimulating discussions.
The work of I.K. was supported in part by RFFI grant 95-01-00755.
He also thanks l'Institut Henri Poincar\'e for the hospitality.

\section*{Appendix A}
\def \theequation {A.\arabic{equation}}
\setcounter{equation} 0

Let us recall the definition of the Riemann theta-function and
some of its properties \ci{D}. We choose the basis of cycles $\alpha_j$,
$\beta_j$ on the Riemann surface $\Gamma_N$ with the canonical matrix of
intersections as shown in Fig.~1. Then, we construct the basis of normalized
holomorphic differentials \re{1st} and evaluate $g\times g$ Riemann matrix
of their $\beta-$periods
\be
\frac1{2\pi}\oint_{\beta_j} d\omega_k=\tau_{jk}
\,,\qquad
\frac1{2\pi}\oint_{\alpha_j} d\omega_k=\delta_{jk}
\,.
\lab{R-B}
\ee
The $\tau-$matrix is symmetric and it has a positive definite imaginary
part. The basic $g-$dimen\-sio\-nal vectors $(e_j)_k=2\pi\delta_{jk}$
and $(b_j)_k=2\pi \tau_{jk}$ generate the lattice $\CL$ and allow
to construct the torus $\CJ(\Gamma_N)=\IC^g/\CL$ called
Jacobian of the Riemann surface $\Gamma_N$.

One defines the $g-$dimen\-sional Riemann theta-function as
\be
\theta(u;\tau)=\sum_{m\in \IZ^g} \exp\lr{i\pi \vev{\tau m,m} + i \vev{u,m}}
\lab{theta}
\ee
where $u=(u_1,...,u_g)$ is a complex vector on the Jacobian
$\CJ(\Gamma_N)$, $m=(m_1,...,m_g)$ is a vector with integer components,
$\vev{\tau m,m}=\sum_{j,k=1}^g \tau_{jk}m_j m_k$ and
$\vev{u,m}=\sum_{j=1}^g u_j m_j$.

The theta-function has the following periodicity properties. For
any integer $l\in \IZ^g$ and complex $u\in \IC^g$ one has
\be
\theta(u+2\pi l)=\theta(u) \,,\qquad
\theta(u+2\pi \tau l)=\exp\lr{-i\pi\vev{\tau l,l}-i\vev{u,l}}\theta(u)\,.
\lab{tran}
\ee

The zeros of the theta-function satisfy the following condition.
Let us fix a reference point $\gamma_0$ on $\Gamma_N$ and define
the vector $A_k(Q)=\int_{\gamma_0}^Q d\omega_k$ for any $Q\in \Gamma_N$.
Then, for an arbitrary complex $Z=(Z_1,...,Z_g)\in \IC^g$ the
function $\theta(A(Q)-Z)$ either vanishes identically or it has
exactly $g$ zeros at the points $Q_1$, $...$, $Q_g$ such that
\be
A(Q_1)+ ... + A(Q_g)=Z-\CK
\lab{zero}
\ee
with $\CK=\CK(\Gamma_N,\gamma_0)$ being the vector of Riemann constants.

\section*{Appendix B}
\def \theequation {B.\arabic{equation}}
\setcounter{equation} 0

Let us establish useful relations between parameters $\Delta_k$ and
$\U_0^{\j}$ defined in
\re{W} and periods of normalized differentials on $\Gamma_N$. Their
derivation is based on the following identity (Stokes theorem) valid
for any two unnormalized differentials $d\omega$ and $d\omega'$ on $\Gamma_N$
\be
\sum_{j=1}^g (a_j b'_j - a_j' b_j)
=  \oint_C d \omega'(Q) \int_{\gamma_0}^Q d \omega
= -\oint_C d \omega(Q) \int_{\gamma_0}^Q d \omega'\,.
\lab{Stokes}
\ee
Here, $a_j=\oint_{\alpha_j} d\omega$, $b_j=\oint_{\beta_j} d\omega$
and similarly for $d\omega'$ are periods of the differentials,
$\gamma_0$ is a point on $\Gamma_N$ in general position, $C$ is an
oriented closed contour on $\Gamma_N$ which encircles all singularities
of differentials. The contour $C$ should not cross the cycles $\alpha_s$
and $\beta_s$ ($s=1,...,g$). The same condition is imposed on the
integration path between points $\gamma_0$ and $Q$ in \re{Stokes}.

Applying the identity \re{Stokes} to a pair of the normalized differentials
$d\omega_k$ and $d\Omega_0$ defined in \re{1st} and \re{Om0} one uses
normalization condition on the l.h.s.\, calculates the r.h.s.\ by taking the
residue at the poles $Q_0^\pm$ and gets the well-known relation \ci{D}
$$
\int_{Q_0^-}^{Q_0^+} d \omega_k = -i \oint_{\beta_k} d\Omega_0 \,.
$$
Let us consider the following unnormalized dipole differential
$$
d\widehat\Omega_0=\frac{2ihdE}{Ey(E)}
              =\frac{2ihdE}{\sqrt{\Lambda^2(E)-4}}\,,
$$
which appears as a leading term in the expansion \re{Om0} of the normalized
dipole differential $d\Omega_0$. It has simple poles at the points
$Q_0^\pm$ with the residue $\pm 1$, respectively, and its asymptotics
at the vicinity of puncture $Q_\infty^+$ on the upper sheet is
\be
d\widehat\Omega_0\stackrel{E\to\infty}{=}
\frac{2ih}{q_N} dE \, E^{-N} \lr{1+\CO(1/E)}\,.
\lab{van}
\ee
Applying the identity \re{Stokes} to a pair of $d\widehat\Omega_0$ and
normalized meromorphic differential of the 2nd kind, $d\Omega^{\j}$, one
takes into account normalization condition \re{3rd-as}, calculates the
r.h.s.\ by taking residue at the simple poles $Q_0^\pm$ of the
differential $d\widehat\Omega_0$ and $j-$th order poles $Q_\infty^\pm$
of the meromorphic differential $d\Omega^{\j}$ and arrives at the
following relation
$$
\sum_{k=1}^{N-2} \oint_{\alpha_k} 
d\widehat\Omega_0 \oint_{\beta_k} d\Omega^{\j}
=-2\pi i \int_{Q_0^-}^{Q_0^+} d\Omega^{\j} + 2\cdot 2\pi i \frac1{(j-1)!}
\lr{\frac{d}{dz}}^{j-1}\frac{d\widehat\Omega_0}{dz}\bigg|_{z=1/E\to 0}\,.
$$
We notice that the derivative in the last term vanishes for
$j=1,...,N-2$ due to \re{van} and one gets the identity
\be
i\U_0^{\j}=
\int_{Q_0^-}^{Q_0^+} d\Omega^{\j}
=-\frac{ih}{\pi}\sum_{k=1}^{N-2} \U^{\j}_k
\oint_{\alpha_k}\frac{dE}{\sqrt{\Lambda^2(E)-4}}\,,
\qquad 1 \le j \le N-2\,,
\lab{id}
\ee
with the vector $\U^{\j}$ defined in \re{A(Q)}.

\bb{99}
\bi{Col}  S.C. Frautschi, {\it Regge poles and S-matrix theory\/},
          New York, W.A. Benjamin, 1963;
\\        V. de Alfaro and T. Regge, {\it Potential scattering},
          Amsterdam, North-Holland, 1965;
\\        P.D.B. Collins, {\it An introduction to Regge theory
          and high energy physics\/}, Cambridge University Press, 1977.
\bi{BKP}  J. Bartels, Nucl. Phys. B175 (1980) 365;
\\        J. Kwiecinski and M. Praszalowicz, Phys. Lett. B94 (180) 413.
\bi{Lip2} L.N. Lipatov, Phys. Lett. B251 (1990) 284;  B309 (1993) 394.
\bi{FK}   L.D. Faddeev and G.P. Korchemsky, [hep-th/9404173];
          Phys. Lett. B 342 (1995) 311.
\bi{Lip}  L.N. Lipatov, JETP Lett. 59 (1994) 596.
\bi{Qua}  G.P. Korchemsky, Nucl. Phys. B462 (1996) 333.
\bi{BFKL} E.A. Kuraev,  L.N. Lipatov and V.S. Fadin,
          Phys. Lett. B60 (1975) 50;
          Sov. Phys. JETP 44 (1976) 443; 45 (1977) 199;
\\        Ya.Ya. Balitsky and L.N. Lipatov, Sov. J. Nucl. Phys. 28 (1978) 822.
\bi{pre}  G.P. Korchemsky, preprints LPTHE-Orsay-96-76 [hep-th/9609123]
          (to appear in Nucl. Phys. B) and
          LPTHE-Orsay-96-89 [hep-ph/9610454] (to appear in the Proceedings
          of ICHEP'96, Warsaw, 25-31 July 1996).
\bi{NMPZ} S.P. Novikov, S.V. Manakov, L.P. Pitaevskii and V.E. Zakharov,
          {\it Theory of Solitons: The Inverse Scattering Method\/},
          Consultants Bureau, New York, 1984;
\\        B. Dubrovin, I. Krichever and S. Novikov,
          {\it Integrable systems - I},
          Sovremennye problemy matematiki (VINITI), Dynamical systems - 4
          (1985) 179;
\\        B.A. Dubrovin, V.B. Matveev and S.P. Novikov,
%         {\it Non-linear equations of Korteweg-de Vries type,
%         finite-zone linear operators and abelian varieties\/},
          Russ. Math. Surv. 31 (1976) 59.
\bi{Kr}   I.M. Krichever, Russ. Math. Surv. 32 (1977) 185;
          Func. Anal. Appl. 14 (1980) 531; 11(1977) 12.
\bi{D}    B.A. Dubrovin, Russ. Math. Surv. 36 (1981) 11.
\bi{F}    L.D. Faddeev, private communication.
\bi{SoV}  E.K. Sklyanin, {\it The quantum Toda chain\/},
          Lecture Notes in Physics, vol.\ 226, Springer, 1985, pp.196--233;
          {\it Functional Bethe ansatz\/}, in ``Integrable
          and superintegrable systems'', ed.\ B.A. Kupershmidt, World
          Scientific, 1990, pp.8--33;
          Progr. Theor. Phys. Suppl. 118 (1995) 35 [solv-int/9504001].
\bi{KBBT} I. Krichever, O. Babelon, E. Billey and M. Talon, Amer. Math.
          Soc. Transl. 170 (1995) 83.
\bi{Bet}  G.P. Korchemsky, Nucl. Phys. B443 (1995) 255.
\bi{QISM} L.A. Takhtajan and L.D. Faddeev, Russ. Math. Survey 34 (1979) 11;
\\        E.K. Sklyanin, L.A. Takhtajan and L.D. Faddeev,
          Theor. Math. Phys. 40 (1980) 688;
\\        V.E. Korepin, N.M. Bogoliubov and A.G. Izergin, {\it Quantum
          inverse scattering method and correlation functions\/},
          Cambridge Univ. Press, 1993.
\bi{Wh}   G.B. Whitham, {\it Linear and Nonlinear Waves\/},
          John Wiley, New York, 1974;
\\        H. Flaschka, M.G. Forest and D.W. McLaughlin, Comm. Pure Appl.
          Math. 33 (1980) 739;
\\        S.Yu. Dobrokhotov and V.P. Maslov, J. Sov. Math. 16 (1981) 1433;
\\        B.A. Dubrovin and S.P. Novikov, Russ. Math. Surv. 44 (1989) 35.
\\        I.M. Krichever, Comm. ath. Phys. 143 (1992) 415;
          Comm. Pure Appl. Math. 47 (1994) 437;
\\        B.A. Dubrovin, Comm. Math. Phys. 145 (1992) 195.
\eb
\end{document}